\documentclass[12pt]{article}

\usepackage{amsmath,amssymb,amsthm,amscd,mathtools, graphicx, mathtext,color,wrapfig}
\usepackage[numbers,sort&compress]{natbib}
\makeatletter

\@addtoreset{equation}{section}
\makeatother
\allowdisplaybreaks[1]
\newcommand{\beq}{\begin{equation}}
\newcommand{\eeq}{\end{equation}}

\theoremstyle{definition}

\begin{document}
\baselineskip=18pt  
\baselineskip 0.713cm

\begin{titlepage}


\renewcommand{\thefootnote}{\fnsymbol{footnote}}

\vskip 1.0cm

\begin{center}
{\LARGE \bf
ADE Little String Theory
\vskip 1cm
on a Riemann Surface
\vskip 1cm
(and Triality)
}

\medskip

\vskip 0.5cm

{\large
Mina Aganagic$^{1,2}$, Nathan Haouzi$^{1}$
}
\\
\medskip

\vskip 0.5cm

{\it
$^1$Center for Theoretical Physics and $^2$Department of Mathematics\\
University of California, Berkeley,  USA\\
}

\end{center}

\vskip 0.5cm

\centerline{{\bf Abstract}}
\medskip
We initiate the study of $(2,0)$ little string theory of ADE type using its definition in terms of IIB string compactified on an ADE singularity. As one application, we derive a 5d ADE quiver gauge theory that describes the little string compactified on a sphere with three full punctures, at low energies. As a second application, we show the partition function of this theory equals the 3-point conformal block of ADE Toda CFT, $q$-deformed. To establish this, we generalize the $A_n$ triality of \cite{AHS} to all ADE Lie algebras; IIB string perspective is crucial for this as well. 
\noindent\end{titlepage}
\setcounter{page}{1} 

\section{Introduction}

String theory has led to remarkable insights into quantum field theories in various dimensions and many areas of mathematics. Recently, a plethora of new conjectures about properties of gauge theories were obtained from studying the six-dimensional $(2,0)$ CFT, whose existence is implied by string and M-theory. The 6d theory seems poised to play an important role in mathematics as well, see e.g. \cite{W7, G2, Gaiotto:2009hg, AGT, Witten:2009at, wittenk}. Yet, the lack of a good description of the $(2,0)$ theory limits what predictions we are able to obtain.

The $(2,0)$ CFT is labeled by an ADE Lie algebra ${\bf g}$ and defined as the limit of IIB string theory on an ADE surface $X$ where we send the string coupling to zero to decouple the bulk modes, and then $m_s$ to infinity to get a CFT. If we keep the string scale finite instead, the result is a six-dimensional string theory, the $(2,0)$ little string theory \cite{Witten:1995zh,Seiberg:1997zk,Losev:1997hx}. The description of the little string theory in terms of IIB string on an ADE singularity in the limit of zero string coupling gives us an essentially complete control of the theory, away from the singularity at the origin of its moduli space \cite{Witten:1995zh}. 

In this paper, we initiate the study of $(2,0)$ little string theory, in connection to gauge theories and mathematics. This perspective is fruitful for the 6d CFT as well: we simply take the string scale to infinity in the very end. For illustration, we will give several applications of this approach. 

We place the $(2,0)$ little string on a Riemann surface ${\cal C}$ which is a cylinder, with co-dimension two defects at points of ${\cal C}$. We identify the defects, using known facts about IIB string on ADE singularities, as D5 branes wrapping the (non-compact) 2-cycles of $X$. We show that the codimension two defects preserving superconformal invariance can be classified using the relation \cite{Reid:1997zy} between the geometry of $X$ and the representation theory of ${\bf g}$. We argue that, except at the very origin of the moduli space, the $(2,0)$ little string theory with defects has a low energy description as a 5d ${\cal N}=1$ ADE quiver theory compactified on a (T-dual) circle. The theory is five-dimensional due to string winding modes around the compact direction in ${\cal C}$. It can be determined by perturbative IIB string analysis \cite{DM} for any collection of defects introduced by D5 branes. As an example, we generalize the gauge theory description of the so called 5d $T_N$ theory, which was recently obtained in the $A$-series case in \cite{AHS, Bergman:2014kza,Hayashi:2014hfa}, to all ADE groups. The gauge theory description of the low energy physics lets us obtain the supersymmetric partition function of the little string on ${\cal C} \times {\mathbb R}^4$, with ${\mathbb R}^4$ regulated by $\Omega$-background, using techniques of \cite{Moore:1997dj, Losev:1997wp,  N2, NO, NP, NPS}. This should be contrasted with the 6d CFT on ${\cal C}$, which generically has no Lagrangian description, so there is no direct way to obtain the partition function. We also give a simple derivation of the Seiberg-Witten curve of the theory by compactifying on an additional circle, and using T-duality to relate D5 branes to monopoles of an ADE gauge theory on ${\mathbb R}\times T^2$, studied in \cite{NP, NPS}.

The second application is to the conjectural relation between the $(2,0)$ 6d CFT and 2d conformal field theory on ${\cal C}$ \cite{AGT,Gaiotto:2014bja,IH, Nt}. We will show that the partition function of the ADE little string on ${\cal C} \times {\mathbb R}^4$ equals the $q$-deformed conformal block of ADE Toda CFT on ${\cal C}$. The $q$-deformed vertex operator insertions are in one to one correspondence with the defect D5 branes. 
The $q$-deformed conformal blocks have a ${\cal W}_{q,t}({\bf g})$ algebra symmetry developed by Frenkel and Reshetikhin \cite{FR1}. They are written
as integrals over positions of screening currents, with integrand which is a correlator in a free theory. 
We show that, computing the integrals by residues, one recovers the partition function of the little string. The numbers of screening charge integrals are related to the values of the Coulomb moduli. In the limit in which we take the string scale to infinity, and little string becomes the $(2,0)$ CFT on ${\cal C}$, the $q$-deformation disappears and one recovers ordinary Toda conformal block, with ${\cal W}({\bf g})$ algebra symmetry.
The relationship between the $(2,0)$ CFT of $A_1$ type and Liouville theory (the $A_1$ Toda CFT) was conjectured by Alday, Gaiotto and Tachikawa, and the correspondence was proven in \cite{Fateev:2009aw, Alba:2010qc}. The $A_n$ version of the correspondence was established in \cite{ACHS, AHS}. The strategy of the proof employed here is the same as in \cite{AHS}, although the string theory realization of $(2,0)$ theory used there is different, related by T-duality. 

We show, in a related application, that there is a {\it triality} of precise relations between three different classes of theories of ADE type: the quiver gauge theories in three and five dimensions and the $q$-deformed Toda CFT. In establishing the correspondence between Toda CFT and the $(2,0)$ theory we described, the central role is played by strings, obtained by wrapping D3 branes on compact 2-cycles in $X$, and at points on ${\cal C}$. The D3 branes are finite tension excitations -- they are vortices on the Higgs branch of little string theory on ${\cal C}$. The theory on D3 branes, derived by a perturbative IIB computation, is a 3d ADE quiver theory compactified on an $S^1$, which in presence of defect D5 branes has ${\cal N}=2$ supersymmetry. The 3d theory turns out to have a manifest relation to Toda CFT: its partition function, expressed as an integral over Coulomb moduli, is identical to the $q$-Toda CFT conformal block -- D3 branes are the screening charges. Interpreting the partition function instead in terms of the Higgs branch of the 3d gauge theory, it equals the 5d partition function, at integer values of Coulomb moduli. Physics of this is the gauge/vortex duality which originates from two different, yet equivalent ways to describe vortices: from the perspective of the theory on the D3 branes or from the perspective of the bulk theory with fluxes. In the latter description, the vortex flux is responsible for shifting the Coulomb moduli by integer values \cite{AHS, ASv} in $\Omega$-background. This is the little string version of large $N$ duality of \cite{DV, DV2, DVp, CKV}. 

The paper is organized as follows. In section 2, we study the ADE $(2,0)$ little string theory on ${\cal C}$ with co-dimension 2 defects, and solve it. In section 3, we study vortices in this theory, in terms of D3 branes, and show how to solve this theory as well. In section 4, we review the free field formulation of ADE Toda CFT, and their $q$-deformation. We show that the $q$-deformed conformal block is the D3 brane partition function. In section 5, we review gauge/vortex duality, and show that the partition function of the ADE $(2,0)$ little string on ${\cal C}$ is the $q$-deformed ADE conformal block on ${\cal C}$. In section 6, we give examples corresponding to the 3-punctured sphere.
We review \cite{FR1} in appendix A.

\section{$(2,0)$ Little String Theory on ${\cal C}$}

ADE little string theory with $(2,0)$ supersymmetry is a six dimensional string theory. It is defined by starting with IIB string theory on an ADE surface $X$, in the limit where we take the string coupling to zero to decouple the bulk modes \cite{Witten:1995zh, Seiberg:1997zk, Losev:1997hx}.\footnote{See \cite{Ahar,Kutasov:2001uf} for review of little string theory. Reviews of many facts about Lie algebras, ADE singularities, quiver gauge theories and monopoles are in \cite{NP}.} 
The surface $X$ is a hyperkahler manifold, obtained by resolving a ${\mathbb C}^2/\Gamma$ singularity where $\Gamma$ a discrete subgroup of $SU(2)$, related to ${\bf g}$ by McKay correspondence. The limit leaves a six-dimensional theory supported near the singularity. The little string theory is not a local quantum field theory. It contains strings whose tension is $m_s^2$ and has a T-duality symmetry that exchanges $(2,0)$ and $(1,1)$ little string, compactified on a circle. The latter is obtained from IIA string on $X$, in the $g_s$ to zero limit. At energies far below the string scale $m_s$, the $(2,0)$ little string reduces to $(2,0)$ 6d conformal field theory, studied in \cite{G2,Gaiotto:2009hg,AGT,Witten:2009at,wittenk} and elsewhere. The little string breaks the conformal invariance of the $(2,0)$ CFT, but it does so in a canonical way.  

The $(2,0)$ theory contains a non-abelian self-dual tensor field based on the Lie algebra ${\bf g}$, but no gravity. 
The moduli space of the little string theory is 
\beq\label{moduli}
{\cal M} = ({\mathbb R}^4 \times S^1)^n/W,
\eeq
where $n$ is the rank, and $W$ the Weyl group, of ${\bf g}$. The scalar fields parameterizing ${\cal M}$ 
come from the moduli of the metric on $X$; they are encoded in the periods of a triplet of self-dual two-forms ${\omega}_{I, J, K}$ and the NS and RR B-fields, along the 2-cycles $S_a$ generating $H_2(X, {\mathbb Z})$.  Their natural normalizations are 
\beq\label{mod}
\int_{S_a} {m_s^4\, \omega_{I,J,K}/g_s}, \qquad \int_{S_a} m_s^2 \, B_{NS}/g_s, \qquad \int_{S_a} m_s^2 \, B_{RR}.
\eeq{}
A power of $g_s$ accompanies NS sector fields but not the RR sector ones, since this is how they enter the low energy action of IIB string. The canonical mass dimension of scalars in a two-form theory is two. In taking $g_s$ to zero we tune the moduli of IIB so that the above combinations are kept fixed. The compact directions in ${\cal M}$ come from periods of the RR B-field and have radius $m_s^2$.  In the low energy limit, when we send $m_s$ to infinity, the moduli space becomes simply $({\mathbb R}^5)^{n}/{W}$, since periodicity of the scalars coming from $B_{RR}$ becomes infinite. The periodicity of scalars coming from $B_{NS}$ would have been $m_s^2/g_s$; it is lost at the outset since $g_s$ is taken to zero. 
 
The perturbative string theory description is good away from the singularity at the origin of the moduli space ${\cal M}$. We are taking $g_s$ to zero, in addition, so from IIB string perspective the theory is under excellent control. To obtain results pertaining to the $(2,0)$ CFT, we will want to take the $m_s$ to infinity limit, but only at the very end.\footnote{The singularity of the world sheet CFT that emerges at the origin of ${\cal M}$ was described in \cite{Witten:1995zh}. We will not need to worry about the breakdown of perturbation theory, as we will stay away from this point.  }

\subsection{Little String on ${\cal C}$ with Codimension Two Defects}
We compactify the $(2,0)$ little sting theory on a Riemann surface ${\cal C}$. Here, we will only consider the simplest possibility, where ${\cal C}$ is a cylinder with flat metric, so that $X\times {\cal C}$ is a solution of IIB string theory. 
We would like to introduce codimension two defects in the little string theory, which are at points on ${\cal C}$ and fill the remaining 4 directions. In the little string limit, IIB string has an essentially unique candidate for such an object: these are D5 branes that wrap {\it non-compact} 2-cycles in $X$, which are points on ${\cal C}$, and fill the rest of the space time.\footnote{String-like defects in 6d SCFTs were given an analogous description in \cite{DelZotto:2015isa}, replacing D5 with D3 branes. This leads to degenerate vertex operators of ${\cal W}_{q,t}(\bf g)$ algebra \cite{AHC}. } The other candidates either have infinite tension as we take $g_s$ to zero, or decouple from the degrees of freedom of the little string. The choice of 2-cycles in $X$ for D5 branes to wrap are governed by symmetries we would like to preserve.  We will choose to preserve half the supersymmetry and conformal invariance of the low energy theory. To translate this into a geometric condition on 2-cycles, we need to first recall some elements of the geometry of ALE spaces \cite{Reid:1997zy, GVW}. Codimension two defects of the $(2,0)$ CFT have been studied by different means in \cite{G2, Wyllard:2009hg, Td, Tachikawa:2009rb, Chacaltana:2012zy}.

\subsubsection{Elements of Holomogy}
\label{subsubsec:elements}

The second homology group $H_2(X, {\mathbb Z})$ of $X$ is a lattice whose generators are $n$ 2-spheres $S_a$, supported at the singularity of $X.$ The 2-spheres intersect according to the Dynkin diagram of ${\bf g}$:
\beq\label{intc}
\#(S_a \cap S_b)= - C_{ab}.
\eeq
The Dynkin matrix $C_{ab}$ is given in terms of the adjacency matrix  $I_{ab}$ of the Dynkin diagram as follows
$$
C_{ab}= 2 \delta_{ab} -I_{ab}.
$$
As a lattice, $H_2(X, {\mathbb Z})$ is the same as the root lattice $\Lambda$ of ${\bf g}$: the homology classes of
cycles $S_a$ correspond to {simple positive roots} $e_a$ of the Lie algebra, the intersection pairing is the inner product on $\Lambda$, up to an overall sign. 

The relative homology group $H_2(X, \partial X, {\mathbb Z})$ of $X$ corresponds \cite{GVW} to the weight lattice $\Lambda_*$ of ${\bf g}$. Here, one allows 2-cycles with boundary at infinity $\partial X$. A two cycle is trivial in $H_2(X, \partial X, {\mathbb Z})$ if it is a boundary of a three-cycle in $X$ up to addition of a 2-cycle at infinity $\partial X$. The group contains the $H_2(X, {\mathbb Z})$ as a sub-lattice, by restricting to compact 2-cycles. Correspondingly, the root lattice is a sub-lattice of the weight lattice $\Lambda\subset \Lambda_*$. The group is spanned by classes of non-compact 2-cycles ${ S}^*_a$, which are, up to sign, the fundamental weights $w_a$ of ${\bf g}$. The fundamental weights are defined by $(e_a, w_b) = \delta_{ab}$, so that 
\beq\label{int}
\#(S_a \cap {S}^*_b) = \delta_{ab}.
\eeq
To construct a cycle in the homology class of $S_b^*$, one starts by zooming in on the neighborhood of the vanishing 2-cycle $S_b$, which is locally $T^*S^2$. Next, one picks a point on $S_b$, away from intersections with other minimal 2-cycles and takes $S_b^*$ to be the fiber of the cotangent bundle to $S_b$ above that point. Per construction, this satisfies \eqref{int}, and extends to the boundary of $X$. 
We distinguish the cycles in $X$ from their homology classes, denoted by $[..]$. For example, the homology class of the $a$-th vanishing 2-cycle $S_a$ is the simple root $[S_a]=e_a$, and a class of the dual non-compact cycle $S_a^*$ is minus the fundamental weight $[S_a^*] = - w_a.$
\begin{figure}[h!]
\emph{}
\hspace{-7ex}
\centering
\vspace{-20pt}
\includegraphics[width=0.5\textwidth]{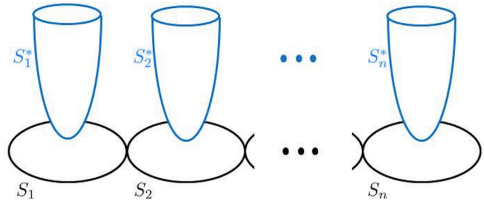}
\vspace{-20pt}
\caption{The vanishing cycles of $A_n$ singularity $S_a$ (in black) and the dual non-compact cycles $S_a^*$ (in blue). For any ADE singularity, $S_a^*$ is constructed as the fiber of the cotangent bundle $T^*S_a$ over a generic point on $S_a$.} 
\label{fig:cycles}
\end{figure}
\subsubsection{Homology Classes of Defects}
\label{subsubsec:classesofdefects}.

To specify a defect D5 brane charge, we pick a collection of non-compact two-cycles $S^*$ 
whose homology class in $H_2(X, \partial X, {\mathbb Z}) = \Lambda_*$ is 
\beq\label{ncomp}
[S^*] = -\sum_{a=1}^n \, m_a \, w_a \;\;  \in \;\; \Lambda_*
\eeq
with positive integer coefficients $m_a$.  A-priori, $m$'s can be arbitrary, but we would like to preserve conformal invariance in 4d, which imposes constraints. A necessary condition for conformal invariance is that the net D5 brane flux vanishes: a non-zero flux of $H_{RR}$ would lead to varying periods of $B_{RR}$ (and running of the gauge coupling constant on the D5 branes).  To satisfy the condition, we add D5 branes that wrap a compact homology class $[S]\in H_2(X, {\mathbb Z})=\Lambda$:
\beq\label{comp}
[S] = \sum_{a=1}^n  \,d_a\,e_a\;\;  \in \;\; \Lambda
\eeq
where $d$'s are also non-negative integers, 
such that
\beq\label{conf}
[S+S_*] =0.
\eeq
In adding \eqref{comp} and \eqref{ncomp}, we used that the root lattice embeds into the weight lattice and the compact homology group into the relative homology group by considering cycles with trivial boundary. The vanishing of $S+S^*$ in homology implies its intersection with any 2-cycle $S_a$ vanishes.  With help of \eqref{intc} and \eqref{int}, we can write this as
\beq\label{conformal}
\sum_{b=1}^n C_{ab} \;d_b = m_a.
\eeq

To preserve supersymmetry, it is not enough to choose the class of $S_*$, we must choose the actual cycles in it. D5 branes wrapping different components of $S_*$ preserve the same supersymmetry if their central charges are aligned. These, in turn, are determined by the periods of the triplet of self-dual two forms ${\vec \omega} = (\omega_I, \omega_J, \omega_k)$ on the non-compact cycles $S_a^*$. Supersymmetry is preserved if they determine a collection of vectors, 
$\int_{S^*_b} \vec{ \omega}$, which point in the same direction for all $b$, corresponding to all the central charges being aligned. Then, all the non-compact D5 branes preserve the same half of supersymmetries of the $(2,0)$ theory. Up to a rotation under which $\vec \omega$ is a vector, we can choose 
$$\int_{{S^*_a}} \omega_I>0, \qquad \int_{{S^*_a}} \omega_{J,K}=0.
$$
Next, we pick a metric on $X$ by picking periods of $\omega_{I,J,K}$ through the compact cycles $S_a$. The choice we make will affect the supersymmetry that D5 branes wrapping compact 2-cycles preserve. It does not affect the non-compact D5 branes, which extend to infinity in $X$, as it only affects the data of $X$ near the singularity. We will begin by setting 
\beq\label{FI}
 \int_{S_a} \omega_{J,K} =0,\;   \int_{S_a} B_{NS}=0, 
\eeq
for all $a$'s and letting
\beq\label{taua}
\tau_a =   \int_{S_a} \, ( m_s^2 \,\omega_I/g_s + i  \, B_{RR} )
\eeq
be arbitrary complex numbers with ${\rm Re}(\tau_a)>0$. Recall that $X$ has a sphere's worth of choices of complex structure. In the complex structure in which $\omega_I$ is a $(1,1)$ form, and having chosen \eqref{FI}, \eqref{taua}, both $[S_b^*]$ and $[S_a]$ have holomorphic 2-cycles representatives, and the D5 branes wrapping both the compact and the non-compact 2-cycles preserve the same supersymmetry. The fact that all the D5 branes preserve the same supersymmetry is important, as it leads to an ADE quiver gauge theory description at low energies, with the quiver diagram based on the Dynkin diagram of ${\bf g}$. 

\subsubsection{Gauge Theory Description of The System}
 \label{subsubsec:5dgaugetheory}
We will now determine the low energy description of the compactified $(2,0)$ little string with defects. For generic $\tau$'s, at energies below the string scale, the entire system can be described in terms of a 5d ${\cal N}=1$ gauge theory which originates from the D5 branes.

At long distances, if $\tau$'s are not zero, the bulk theory is a theory of abelian self-dual 2-forms. The 2-forms are non-dynamical from the perspective of the compactified theory, since they propagate in all six dimensions.  At the same time, $\tau$'s determine the inverse gauge couplings of the D5 brane gauge theory. As long as they are non-zero, the theory on the D5 branes has a gauge theory description at low energies.\footnote{The 
 $1/g_{YM}^2$ in five dimensions has units of mass. The $\tau$ is the dimensionless combination $\tau \sim 1/(g_{YM}^2 m_s)$.  The gauge theory description is applicable for energies $E/m_s$ less than $1$, and the theory is weakly coupled for $E/m_s$ less than $\tau$. When we study the partition function, $\exp(-\tau)$ will be the instanton expansion parameter, and we will want this to be less than $1$, so we only need ${\rm Re}(\tau)>0$.} Thus, for non-zero $\tau$ and below the string scale, the dynamics of the $(2,0)$ little string theory on ${\cal C}$ with defects can be described by the gauge theory {\it on the D5 branes}.
 
String theory allows us to determine the gauge theory on the D5 branes on $X$.  
 The gauge theory on the D5 branes wrapping $S$ was worked out in \cite{DM}.
It is an ADE quiver gauge theory, with gauge group
\beq\label{5dgaugegroup}
\prod_{a=1}^n U(d_a),
\eeq
and $I_{ab}$ hypermultiplets in the bi-fundamental $(d_a, \overline{d_b})$ representation for each pair of nodes $a$ and $b$. The theory has ${\cal N}=2$ supersymmetry in four dimensions, since D5 branes break half the supersymmetry of IIB on $X.$ The rank $d_a$ of the gauge group associated to the $a$-th node of the quiver is the number of D5 branes wrapping the 2-cycle ${S_a}$. The hypermultiplets come from the intersections of cycles $S_a$ with $S_b$. This follows from a computation we can do locally, near an intersection point.
A non-zero intersection number $I_{ab}$ of $S_a$ with $S_b$, for distinct $a$ and $b$, means that they intersect transversally at $I_{ab}$ points. At a transversal intersection of 2 holomorphic 2-cycles in a 4-manifold, there are 4 directions in which open strings with endpoints on the branes have DN boundary conditions, leading to a massless bi-fundamental hypermultiplet. The $U(1)$ gauge groups of the D5 branes wrapping cycles are actually massive, by Green-Schwarz mechanism \cite{DM}, so the gauge groups are $SU(d_a)$ not $U(d_a)$. Correspondingly, the Coulomb moduli associated with the $U(1)$ centers are parameters of the theory, not moduli.   Nevertheless, the effects of these $U(1)$'s remain: for example, due to stringy effects  \cite{Losev:2003py}, the partition function is that of a $U(d_a)$ theory. For this reason, we will write the gauge group with $U(1)$ factors included, trusting the reader can keep in mind the subtle point. (The issue of the $U(1)$'s was discussed in \cite{Bergman:2014kza,Hayashi:2014hfa}, from a related perspective.) The D5 branes on $S^*$ do not contribute to the gauge group, since the cycle is non-compact, but they do contribute matter fields: The intersections of non-compact cycles with compact cycles lead to  additional fundamental matter hypermultiplets. Since $S^*_a$ correspond to fundamental weights $w_a$, they do not intersect $S_b$ for $b\neq a$, see \eqref{int}. Thus, with $S^*$ as in \eqref{ncomp}, there are 
$m_a$ fundamental hypermultiplets on the $a$'th node. In section \ref{sec:Examples}, we will work out examples of 5d quiver gauge theories that describe the corresponding little string theory on a sphere with three full punctures. The resulting quivers are given in figures 2-6. In the $A_n$ case, this corresponds to the so called $T_{N}$ theory, with $N=n+1$; this quiver was obtained earlier in \cite{AHS} and studied further in \cite{Bergman:2014kza,Hayashi:2014hfa}.  The rest are new.%
\begin{figure}[h!]
\emph{}
\hspace{-8ex}
\centering
\vspace{-35pt}
\includegraphics[width=0.5\textwidth]{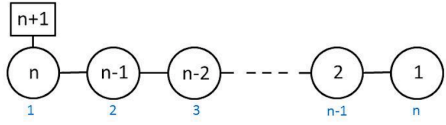}
\vspace{-20pt}
\caption{5d gauge theory describing $A_n$ little string with 3 full punctures.} 
\vspace{-20pt}
\label{fig:Anfigure}
\end{figure}
\begin{figure}[h!]
\emph{}
\hspace{-7ex}
\centering
\vspace{-20pt}
\includegraphics[width=0.5\textwidth]{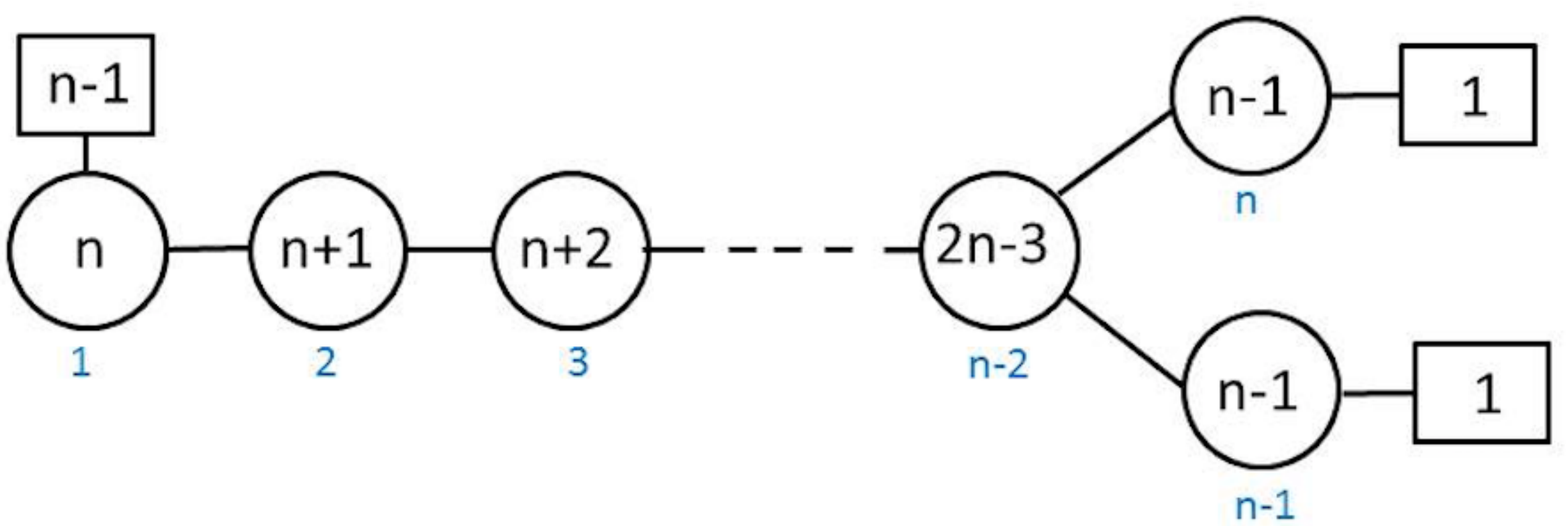}
\vspace{-20pt}
\caption{5d gauge theory describing $D_n$ little string with 3 full punctures.} 
\label{fig:Dnfigure}
\end{figure}
\begin{figure}[h]
\emph{}
\hspace{-7ex}
\centering
\vspace{-20pt}
\includegraphics[width=0.55\textwidth]{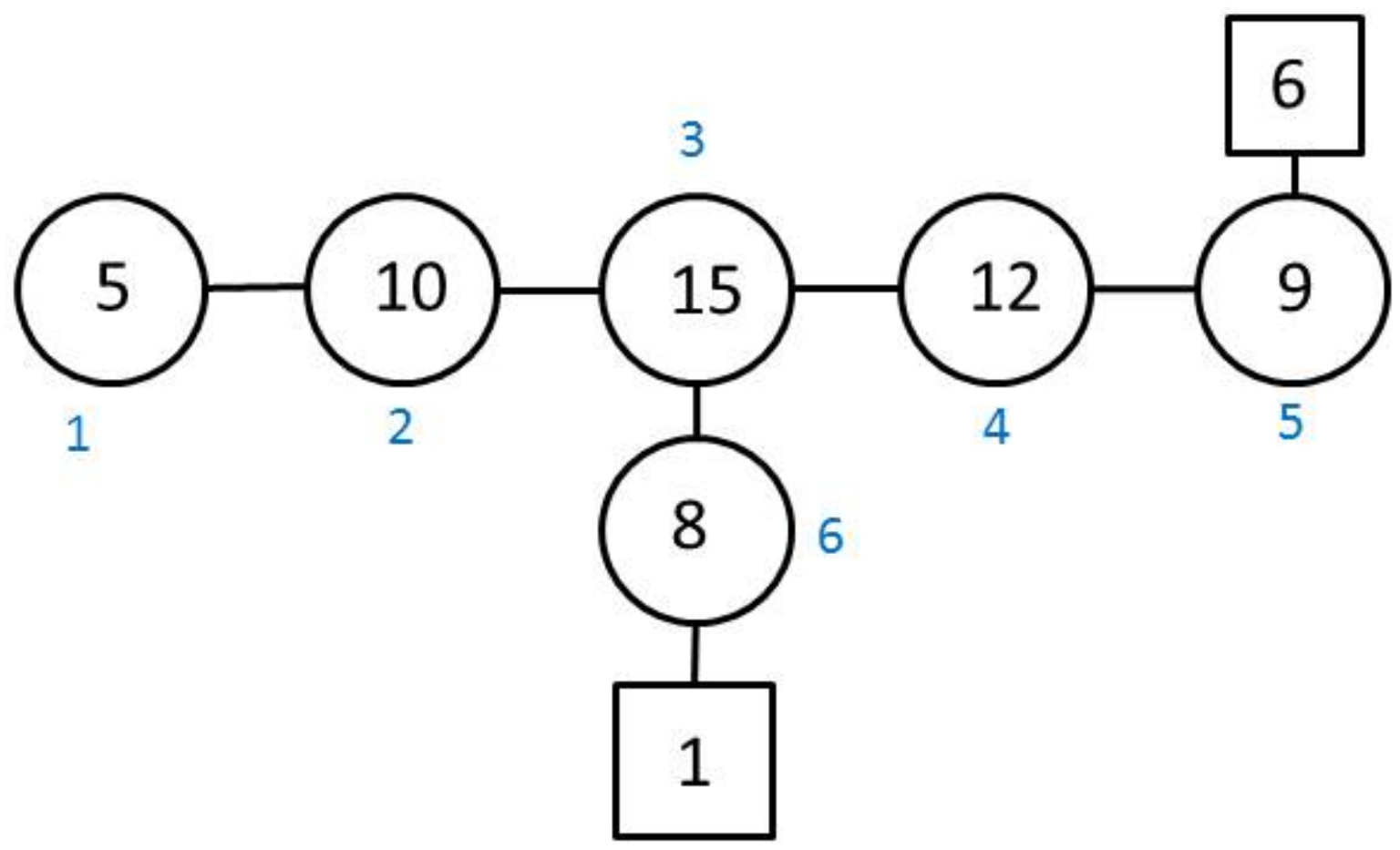}
\vspace{-20pt}
\caption{5d gauge theory describing $E_6$ little string with 3 full punctures.}
\label{fig:E6figure}
\end{figure}
\begin{figure}[h!]
\emph{}
\hspace{-7ex}
\centering
\vspace{-20pt}
\includegraphics[width=0.55\textwidth]{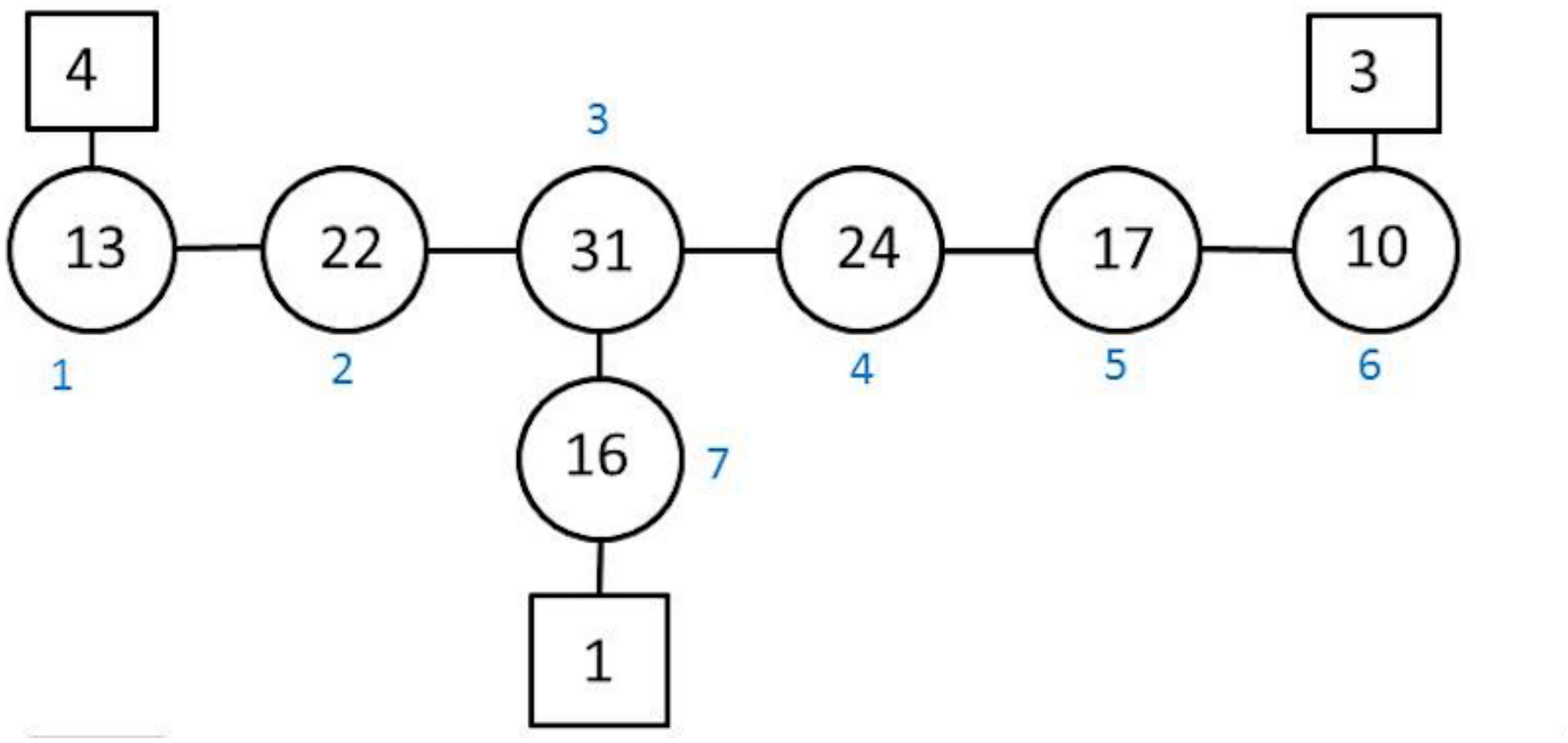}
\vspace{-20pt}
\caption{5d gauge theory describing $E_7$ little string with 3 full punctures.}
\vspace{-20pt}
\label{fig:E7figure}
\end{figure}
\begin{figure}[h!]
\emph{}
\hspace{-7ex}
\centering
\vspace{-20pt}
\includegraphics[width=0.55\textwidth]{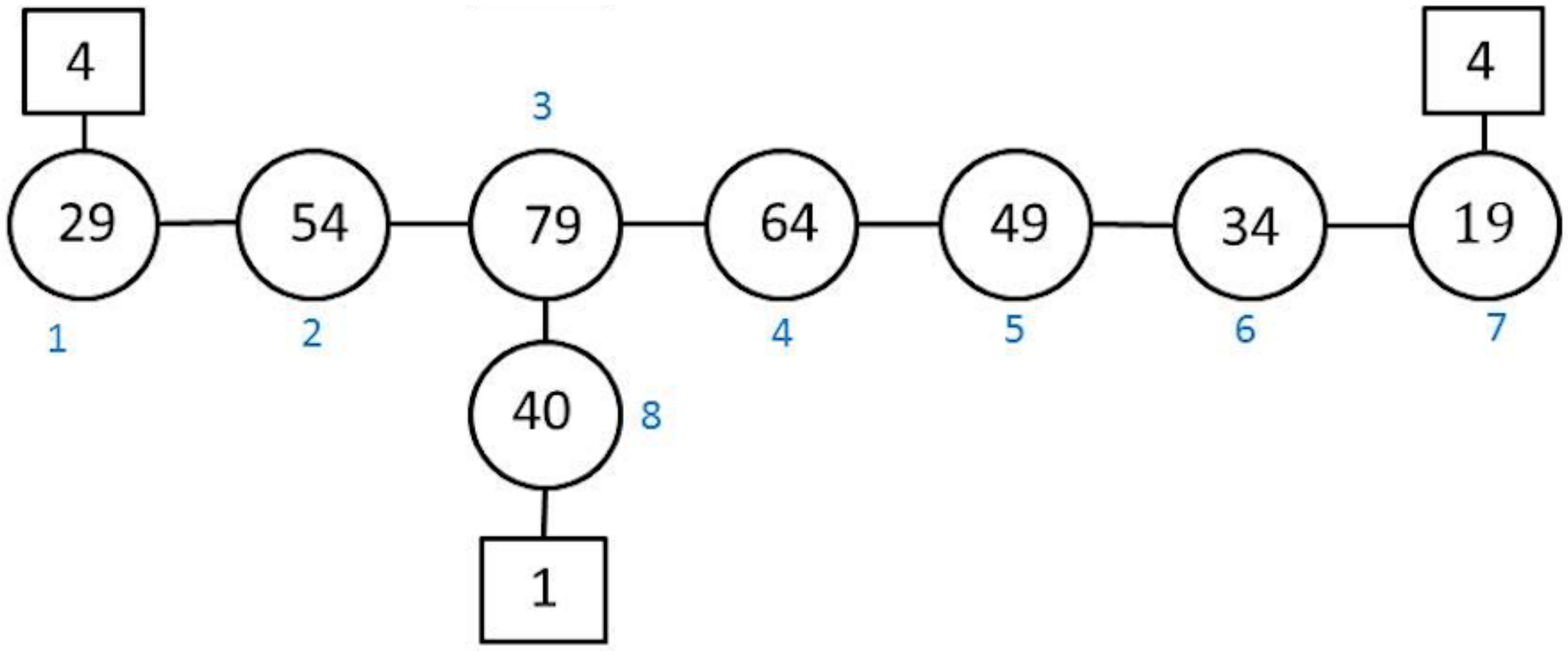}
\vspace{-20pt}
\caption{5d gauge theory describing $E_8$ little string with 3 full punctures.}
\label{fig:E8figure}
\end{figure}

While the theory has the super-Poincare invariance of a 4d ${\cal N}=2$ theory, it is a 5d ${\cal N}=1$ theory compactified on a circle of radius $R$.  Recall that D5 branes are points on ${\cal C}={\mathbb R} \times S^1({\hat R})$, and we are keeping $m_s$ finite.
The zero modes of strings that wind around $S^1({\hat R})$ lead to a Kaluza-Klein tower of states on the T-dual circle of radius 
\beq\label{radius}
R = {1\over m_s^2 { \hat R}}.
\eeq
The resulting tower of states affects the low energy physics \cite{Lawrence:1997jr}. For example, the supersymmetric partition function of the theory depends on $R$, as we will review in section \ref{subsec:5dpart}. (Another way to see this it to do T-duality on the circle. This relates D5 branes which are points on $S^1({\hat R})$ to D6 branes wrapping $S^1(R)$. In the D6 brane description, the fact that the low energy theory is a five dimensional theory on a circle of radius $R$ is manifest.)  

The moduli of the $(2,0)$ theory in six dimensions become parameters in four dimensions.
They determine the couplings of the D5 brane gauge theory. The complex combinations of the moduli which we called $\tau_a =  \int_{S_a} ({m_s^2\over g_s}\, \omega_I+ i   B_{RR})$ are the gauge couplings of the effective 4d gauge theory on the D5 branes. The triplets of ${\cal N}=2$ Fayet-Iliopolous parameters, one for each node of the quiver, come from the remaining 6d moduli, $\int_{S_a} {m_s^2\, \omega_{J,K}/g_s} $ and $\int_{S_a} B_{NS}/g_s$. The only other parameters in the theory are the masses of the fundamental hypermultiplets. These come from the positions of the non-compact D5 branes on ${\cal C}$: the non-compactness of the cycles in $X$ renders these non-dynamical as well. Finally, since the $U(1)$ centers of the gauge group are not dynamical, the Coulomb moduli associated with them are parameters of the theory as well. 

\subsubsection{$(2,0)$ CFT From The Little String}\label{subsubsec:CFTl}

The $(2,0)$ little string theory on ${\cal C}$ is not conformal. To recover the $(2,0)$ 6d CFT theory on ${\cal C}$ we need to take the string scale $m_s$ to infinity\footnote{More precisely, we are considering energies $E$ where $E/m_s$ goes to zero, and $E {\hat R}$ is finite. In the limit, $ER = E/(m_s^2{\hat R})$ goes to zero, so the theory becomes four dimensional. Keeping the $(2,0)$ modulus $\phi$ fixed means $\phi/E^2$ is finite in the limit. If $\phi= \tau m_s^2$, where $\tau$ is the D5 gauge coupling, then $\tau$ must go to zero. This is the strong coupling limit.}, while keeping the Riemann surface ${\cal C}$ and the moduli of the $(2,0)$ theory in \eqref{mod} fixed. Furthermore, we will send to zero $\Delta x m_s$, where $\Delta x$ are the relative positions of D5 branes on ${\cal C}$.

The gauge theory on D5 branes becomes four-dimensional since the radius $R = 1/(m_s^2 {\hat R}) $ of the 5d circle vanishes in the limit. The Lagrangian description of it from the previous section breaks down: the inverse gauge coupling $\tau_a$ of the 4d theory vanishes since $m_s^2 \tau$ is one of the moduli of the $(2,0)$ SCFT, and needs to stay fixed in the limit. There is no energy scale where the ADE quiver gauge theory description is weakly coupled. As we will see, for ${\bf g}\neq A_n$, the effective rank of the gauge group becomes smaller in the limit. Nevertheless, we can learn a lot about the CFT by working with the mass-deformed theory, and taking the massless limit only at the very end. As is commonly the case, the massive theory is easier to understand. 

Finally, note that there is another way to take the $R$ to zero limit, where we keep $\tau$'s finite. This would result in a 4d conformal field theory with the same quiver as the 5d theory, studied in \cite{NP, NPS}. This does not describe the $(2,0)$ theory on ${\cal C}$, as the moduli of the $(2,0)$ theory, proportional to $\tau_a m_s^2$, go to infinity in the field space.

\subsection{Partition Function of Little String Theory on ${\cal C}$}
 \label{subsec:5dpart}
We can use the gauge theory description from section \ref{subsubsec:5dgaugetheory} to compute the supersymmetric partition function of the $(2,0)$ little string theory on ${\cal C} \times {\mathbb C}^2$ with an arbitrary collection of defects at points of ${\cal C}$: 
$$
Z_{Little\; String} ({\cal C} \times {\mathbb C}^2)= Z_{5d}\,(S^1\times {\mathbb C}^2)
$$
The supersymmetric partition function should be an RG invariant, so the description from section \ref{subsubsec:5dgaugetheory} is valid at long distances and   should suffice for non-zero $\tau_a$'s.
More precisely,  we will replace ${\mathbb C}^2$ with the four dimensional $\Omega$-background \cite{N2, NO} as a regulator, and then 
the partition function is the trace
$$
 Z_{5d}\,(S^1\times {\mathbb C}^2)  = {\rm Tr} (-1)^F \; g.
$$
Here $g=q^{S_1 - S_R} t^{-S_2 + S_R}$; $S_1$, $S_2$ are generators of the rotations around the two complex planes in ${\mathbb C}^2$, $F$ is the fermion number, and $q=e^{R \epsilon_1}$, $t= e^{-R \epsilon_2}$. To form a supersymmetric trace, we make use of the $SU(2)_R$ R-symmetry that is preserved by the configuration of D5 branes (this is the subgroup of the $SO(5)_R$ R-symmetry of IIB string theory on $X$ which acts by rotating triplets of scalars in \eqref{FI} that vanish in the brane background). We will make use only of the $U(1)_R\subset  SU(2)_R$ subgroup, generated by $S_R$, and which acts by rotations of the scalars coming from $\omega_{K}$ and $B_{NS}$ in \eqref{FI}.  The trace is the trace in going around the 5d circle.
The gauge theory partition function, in addition to $q$ and $t$, depends on $\tau$'s, the 5d gauge couplings, which are the moduli of the $(2,0)$ theory in 6d, see \eqref{taua}.  It depends on the masses of the fundamental hypermultiplets - these are the positions of the non-compact D5 branes on ${\cal C}$. Finally, it depends on the Coulomb moduli; these are the positions of compact D5 branes on ${\cal C}$.

The partition function of quiver gauge theories of this type was computed in \cite{NP}, using equivariant integration on the instanton moduli space. One works equivariantly with respect to all the $U(1)$ symmetries in the problem: the $U(1)$ symmetries coming from the Cartan subalgebra of the D5 gauge and flavor groups, and the rotations $S_1$, $S_2$, and $S_R$. The partition function becomes a sum over fixed points in instanton moduli spaces. The latter are labeled by a
collection $\{R\}$ of 2d partitions:
$$
\{R\} = \{{R}_{(a),i}\}_{ a=1, \ldots n; \, i=1, \ldots d_{a}},
$$
one for each $U(1)$ factor in the gauge group on the D5 branes. The 2d partitions describe how instantons of the corresponding $U(1)$ factor get "stacked" at the fixed point in ${\mathbb C}^2$. At each node, there are as many 2d Young diagrams as the rank of the corresponding unitary gauge group.
The contribution of each fixed point to the partition function,
\beq\label{bN}
{ Z}_{5d} = r_{5d} \;\sum_{\{R\}} \;I_{ 5d,\{R\}}(q,t; a, m, \tau),
\eeq
 is a product of factors
\beq\label{5dp}
I_{5d, \{R\}}= \; e^{\tau \cdot R} \;\; \cdot \prod_{a=1}^n z_{V_a, {\vec R}^{(a)}} \; z_{H_a, {\vec R}^{(a)}}\; z_{CS, {\vec R}^{(a)}} \; \cdot \prod^n_{a,b=1}
z_{H^{a,b}, {\vec R}^{(a)},  {\vec R}^{(b)}},
\eeq
which we can read off from the D5 brane quiver (we follow the conventions of \cite{AWATA:2008ed}).
The gauge group on the $a$-th node of the ADE quiver is $U(d_a)$. The corresponding vector multiplets contribute

$$
z_{V_a, {\vec R}^{(a)}}= \prod_{1\leq I,J\leq d_{a}}[N_{R_{(a),i} R_{(a),j}}(e_{(a),I}/e_{(a),J})]^{-1}.
$$
Here, $e_{(b),I} = \exp(R a_{(b),I})$ encode the $d_a$ Coulomb branch parameters of the $U(d_a)$ gauge group. 
There are $m_a$ hypermultiplets charged in fundamental representation of the $U(d_a)$ gauge group. They contribute to the partition function as:

$$
z_{H_a, {\vec R}^{(a)}} = \prod_{1\leq \alpha \leq m_a} \prod_{1\leq I\leq d_a} N_{\varnothing R_{(a),I}}( v f_{(a), \alpha}/e_{(a), I}).
$$
The masses of the hypermultiplets are encoded in $f_{(a), \alpha}= \exp(R m_{(a), i})$, where $\alpha$ takes $m_a$ values. In what follows, we write $v = {(q/t)^{1/2}}$. 
For every pair of nodes $a,b$ connected by an edge in the Dynkin diagram, we get a bifundamental hypermultiplet. Its contribution to the partition function is:
$$
z_{H^{a,b}, {\vec R}^{(a)},  {\vec R}^{(b)}}= \prod_{1\leq i\leq d_{a}}\prod_{1\leq j\leq d_{b}}[N_{R_{(a),I} R_{(b),J}}(e_{(a),I}/e_{(b),J})]^{I_{a,b}}.
$$
where $I_{a,b}$ is the incidence matrix, equal to either $1$ or $0$, depending on whether, in the Dynkin graph, there is an arrow starting on the $a$'th node and ending on the $b$'th one or not. The contribution of 5d ${\cal N}=1$ Chern-Simons terms $k^{CS}_{a}$ for this node reads
$$
z_{CS, {\vec R}^{(a)}} = \prod\limits_{1\leq I\leq d_a} \big(T_{R_{(a),I}}\big)^{k^{CS}_a}
$$
Here, $T_{R}$ is defined as $ T_R =(-1)^{|R|} q^{\Arrowvert R\Arrowvert/2}t^{-\Arrowvert R^t\Arrowvert/2}$. The 5d ${\cal N}=1$ Chern-Simons terms can be determined by conformal invariance; with the rest of the partition function as written, $k^{CS}_{a}$ on the $a$-the node is the difference of the ranks of the gauge group on that node, and the following node(s).
The gauge couplings keep track of the total instanton charge, via the combination

$$\tau \cdot R = \sum_{a=1}^n  \sum_{I=1}^{d_a}\;\tau_a\; |R_{(a),I}|.
$$
The vector and hypermultiplet contributions are all given in terms of the Nekrasov function  $N_{RP}(Q)$, which is defined as:
\beq\label{nekrasovN}
N_{RP}(Q) = \prod\limits_{i = 1}^{\infty} \prod\limits_{j = 1}^{\infty}
\dfrac{\varphi_q\big( Q q^{R_i-P_j} t^{j - i + 1} \big)}{\varphi_q\big( Q q^{R_i-P_j} t^{j - i} \big)} \
\dfrac{\varphi_q\big( Q t^{j - i} \big)}{\varphi_q\big( Q t^{j - i + 1} \big)}.
\eeq
where $\varphi_q(x) = \prod\limits_{n=0}^{\infty}(1-q^n x)$ is the quantum dilogarithm. 
The normalization factor $r_{5d}$ in \eqref{bN} contains the tree level and the one loop contributions to the partition function.

\subsection{Integrable Systems, Bogomolny and Hitchin Equations}
 \label{subsec:integrable}

The integrable system associated to the $(2,0)$ little string theory on ${\cal C}$ has two descriptions. The first is in terms of the
moduli space of ${G}$ monopoles on ${\mathbb R} \times T^2$.  The second is in terms of a Hitchin-type system on ${\cal C}$. These integrable systems and their relation to 5d quiver gauge theories were studied recently in \cite{NP, NPS, V}, so we can focus here on the new aspect, namely, the relation to the $(2,0)$ little string theory.  
The connection to integrable systems emerges upon compactifying the theory on an additional circle, which we take to have the radius ${\hat R}'$, so we study $(2,0)$ little string on ${\cal C} \times S^1({\hat R'})$, with defects at points on ${\cal C}$, as before. 

\subsubsection{Monopoles on ${\mathbb R}\times T^2$}

The relation to monopoles on ${\mathbb R}\times T^2$ emerges upon T-duality on $S^1({\hat R'})$. T-duality relates IIB 
to type IIA string on  
$X \times {\cal C}\times S^1({R'})$, where ${R'} = 1/(\hat{R' }m_s^2)$. It also relates D5 branes to D4 branes and $(2,0)$ little string on ${\cal C}\times S^1({\hat R'})$ to $(1,1)$ little string on ${\cal C}\times S^1({R'})$, since it remains a symmetry of the theory as long as $m_s$ is finite.
The $(1,1)$ little string becomes, at low energies, the maximally supersymmetric gauge theory in 6d, with gauge group based on the Lie algebra ${\bf g}$. The D4 branes are ${\bf g}$ monopoles: they are points on ${\cal C} \times S^1({R'})$, magnetically charged under the gauge fields of the 6d little string (the gauge fields originate from periods of the RR 3-form potential on the 2-cycles in $X$). The identification of D4 branes with monopoles identifies the Coulomb branch of the D5 brane gauge theory on $S^1({\hat R'})$ with the moduli space of ${\bf g}$-monopoles on ${\cal C} \times S^1({R'})$; the latter is a hyperkahler manifold ${\cal M}_{C}$ of quaternionic dimension $\sum_{a=1}^n d_a$.

\vskip 0.5 cm
{\it D-Branes as Monopoles}
 \vskip 0.5 cm
The D4 branes wrapping the compact 2-cycles are non-abelian monopoles. In gauge theory, the non-abelian monopole charges are valued \cite{NP} in the co-root lattice $\Lambda^{\vee}$ of ${\bf g}$.
The charges of D4 branes wrapping compact 2-cycles live in $H_{cmpt}^2(X, {\mathbb Z})$, corresponding to compactly supported cohomology. This group is indeed the same as the co-root lattice, $ \Lambda^{\vee}=H_{cmpt}^2(X, {\mathbb Z})$ \cite{GVW}. Using Poincare duality, or the inner product on the Lie algebra, we can identify this lattice with the homology of compact support $H_2(X, {\mathbb Z}) = \Lambda$, or equivalently, with the root lattice. This identifies the charge of the D4 brane 
wrapping the cycle $S$ with the  homology class of the cycle $[S] = \sum_{a=1}^n d_a e_a$ itself.


The D4 branes wrapping non-compact cycles, are singular, Dirac monopoles \cite{Cherkis:1997aa,Cherkis:1998hi}. In gauge theory, the charges of Dirac monopoles are supported in the co-weight lattice $\Lambda_*^{\vee}$ of ${\bf g}$. The charges of non-compact branes are in $H^2(X, {\mathbb Z})$, containing cohomology of both compact and non-compact support. The cohomology group $H^2(X, {\mathbb Z}) = \Lambda_*^{\vee}$ is the co-weight lattice, as it is dual to the root lattice $H_2(X, {\mathbb Z}) =\Lambda$.  Poincare duality, in turn, provides identification between $H^2(X, {\mathbb Z})$ with $H_2(X, \partial X; {\mathbb Z})$, so identifies the co-weight and weight lattices, where the identification becomes the identity map for a simply laced ${\bf g}$. The charge of the Dirac monopole corresponding to  D4 branes wrapping the non-compact cycle $S^*$ equals  $[S^*]= -\sum_{a} m_a \, w_a$, the class of the cycle in \eqref{ncomp}.



Monopoles are solutions to Bogomolny equations: 
\beq\label{Bogomol}
D\phi = * F.
\eeq
Here $F$ is the  curvature of the gauge field, and $\phi$ is the real scalar field which approaches a constant value at infinity; both are valued in the Lie algebra ${\bf g}$. In our case $F$ and $\phi$ come from the $(1,1)$ little string. The scalar $\phi$ is $\phi_a =  \int_{S_a} {m_s^3\, \omega^I/ {g_s'}}$, where ${g_s'} $ is the IIA string coupling, related to the IIB string coupling by $m_s/g_s' = \hat{R'} \,m_s^2 /{g_s}$.
We need to solve Bogomolny equations on ${\mathbb R} \times T^2= {\cal C} \times S^1({R'})$. 
 
\subsubsection{Hitchin-Type System on ${\cal C}$}
\label{subsubsec:Hitchintype}
If ${R'}$ is small, we can forget about the positions of the monopoles on the circle $S^1({{R}'})$. This corresponds to going back to the original D5 brane description on a large circle, of radius ${\hat R}'$. Then, the Bogomolny equations on ${\cal C} \times S^1({{R'}})$ reduce to Hitchin-type equations on ${\cal C}$,
\beq\label{Hitch}
\begin{aligned}
F &= \,  [\varphi, \overline \varphi]\\
{\bar D_x} \varphi & = 0 = D_x {\overline \varphi}.
\end{aligned}
\eeq
These are not exactly the equations considered by Hitchin, because the imaginary part of $\varphi$ is a periodic scalar, with period $1/{R}'$. In going  from D4 branes back to D5 branes,  $\phi$  in \eqref{Bogomol} was complexified to $\varphi = \phi + i  A_{\theta}$, where $A_{\theta}$ is the holonomy of the $(1,1)$ gauge field around the $S^1({R}')$ circle. The periodicity of the Higgs field ${\varphi}$ reflects the fact we study 5d theory on a circle, as opposed to a purely four dimensional one. Given a solution for $\varphi(x)$ of \eqref{Hitch} corresponding to D5 brane defects, we can write the Seiberg-Witten curve as the spectral curve of $\varphi$, taken in some representation $Q$ of ${\bf g}$:

\beq\label{5dSWcurve}
{\rm det}_{Q} (e^{R' p} - e^{{ R}' \varphi(x)})=0.
\eeq

The factor of the radius ${ R}'$ accompanying $\varphi$ in the formula is determined so that ${{R}'  \varphi}$ has period $2\pi i$, and the exponent is single valued. One can phrase this as studying a group-valued Hitchin system on ${\cal C}$, see  \cite{V}. (In the language of IIB on $X$, the periodic direction comes from the period of the RR B-field on 2-cycles of ALE.  It is related to the holonomy of the bulk $(1,1)$ 6d gauge field on the circle: $R' A^a_{\theta} = \int_{S^a} B_{RR}$.) We can take the determinant in any representation $Q$ of the group.  

When we send $m_s$ to infinity, the $(2,0)$ little string theory becomes the $(2,0)$ 6d CFT. We want to do this while keeping the scalar fields of the $(2,0)$ theory fixed, and also the radii of the two circles ${\hat R}$ and ${\hat R}'$ it is compactified on. Then the radius ${R'}$ goes to zero, and the 6d $(1,1)$ little string theory compactified on ${\cal C}\times S^1(R')$ becomes 5d, maximally supersymmetric Yang-Mills on ${\cal C}$, with inverse gauge coupling squared equal to  $m_s^2{R'} = 1/{ {\hat R'}}$. At the same time, the periodic direction in $\varphi$ decompactifies, and we recover the ordinary Hitchin system on ${\cal C}$, associated to the Lie algebra ${\bf g}$. This is the integrable system associated to class $S$ theory on ${\cal C}$ as in \cite{Gaiotto:2009hg}. In particular, in the limit, we will recover the spectral curve of the Hitchin integrable system:
\beq\label{4dSWcurve}
{\rm det}_{Q} (p- \varphi(x))=0.
\eeq
 
\subsection{The Weight System ${\cal W}_{\cal S}$}
\label{subsubsec:WS}

It turns out to be very fruitful to study the theory on the Higgs branch, where the gauge group $\prod_{a=1}^n U(d_a)$ is broken to its $U(1)$ centers, one for each node. 
We force the theory onto the Higgs branch by turning on the remaining moduli of the $(2,0)$ theory $ \int_{S_a} \omega_{J,K}, \int_{S_a} B_{NS}$ (see \cite{Xie:2014pua} for a detailed analysis from gauge theory perspective); these are the FI parameters in the D5 brane gauge theory \eqref{FI}. The deformation is normalizable, affecting only the geometry of $X$ near the singularity. 

On the Higgs branch, the compact and non-compact D5 branes must recombine: the deformation changes the supersymmetries preserved by the compact D5 branes (it changes their central charges via \eqref{FI}), but not the supersymmetries preserved by the non-compact ones (these can be detected at infinity in $X$). As a consequence, the branes on $S$ in \eqref{comp} and on $S^*$ in \eqref{ncomp} are no longer mutually supersymmetric. From the monopole perspective, this is the statement that once the gauge group is Higgsed, there are no non-abelian monopole solutions to \eqref{Bogomol}. Instead, all monopoles recombine into Dirac monopoles of the leftover, abelian gauge group. Correspondingly, after Higgsing of the gauge group, the D5 branes wrapping $S$ and $S^*$ re-combine to form D5 branes wrapping a collection of non-compact cycle $S_i^*$, whose homology classes are elements $\omega_i$ of the weight lattice $\Lambda^* = H_2(X, \partial X, {\mathbb Z})$:
\beq\label{weightsfr}
\omega_i  = [S_i^*] \qquad \in  \qquad \Lambda^*.
\eeq
The classes $\omega_i$ can be determined as follows. Each of $\omega_i$'s comes from one of the non-compact D5 branes on $S^*$.
For the branes to bind, the positions of compact branes must coincide with positions of one of the non-compact D5 branes on ${\cal C}$. The positions of non-compact D5 branes are mass parameters of the quiver gauge theory, the positions of compact D5 branes on ${\cal C}$ are Coulomb moduli; when a Coulomb modulus coincides with one of the masses, the corresponding fundamental hypermultiplet becomes massless and can get expectation values. This, in turn, describes the D5 branes binding (see \cite{Kachru:1999vj} for a similar example), and allows supersymmetry to be preserved in presence of non-zero FI terms. 
Thus, $\omega_i$ has the form $ -w_a$ plus the sum of positive simple roots $e_a$, from bound compact branes. Not any such combination will correspond to truly bound branes: a sufficient condition is that  $\omega_i $ is in the Weyl orbit of $-w_a = [S_a^*]$: this means that there is a point in the moduli space of the theory on $X$ where $S_i^*$ and $S_a^*$ look the same. Furthermore, the collection of weights 
\beq\label{WS}
{\cal W}_{\cal S} = \{ \omega_i\}_i
\eeq
we get must be such that it accounts for all the D5 brane charge in $[S^*]$ and in $[S]$. One simple consequence is that the number of $\omega_i$'s is thus the rank of the 5d flavor group, $\sum_{a=1}^n m_a$. The fact that the net D5 charge is zero $[S+S^*]=0$ implies that
\beq\label{sumtozero}
\sum_{\omega_i \in {\cal W}_{\cal S}} \omega_i=0,
\eeq
which is thus equivalent to \eqref{conformal}.
\begin{figure}[h!]
\emph{}
\hspace{-7ex}
\centering
\vspace{-30pt}
\includegraphics[width=0.55\textwidth]{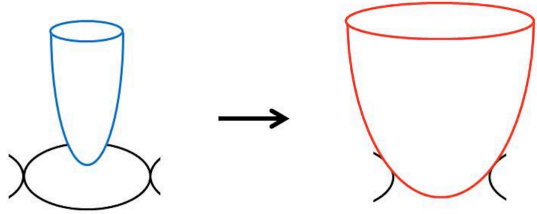}
\vspace{-20pt}
\caption{D5 branes in classes $e_a$ and $-w_a$ bind to a brane in class $-w_a+e_a$.}
\label{fig:WS}
\end{figure}

In section \ref{sec:Examples}, we will work out some simple examples. The most canonical type of defect uses the fact that the weight lattice of a Lie algebra of rank 
$n$ is $n$ dimensional. We can construct a set ${\cal W}_{\cal S}$ by picking any $n+1$ weight vectors which lie in weyl-group orbit of the fundamental weights $-w_a$, for some $a$, and which sum up to zero, and such that $n$ of them span $\Lambda_*$. This leads to a
a full puncture defect of the $(2,0)$ little string on ${\cal C}$, and the corresponding gauge theory description in figures 2-6. We can also string sets of such defects, for a $k$ punctured sphere.

\vskip 0.5cm
{\it The ${\cal S}$-curve from Dirac Monopoles and ${\cal W}_{\cal S}$}
\vskip 0.5cm

The Seiberg-Witten curve \eqref{5dSWcurve},\eqref{4dSWcurve}, captures the low energy physics of the theory on the Coulomb branch and coincides with the spectral curve of the corresponding integrable system. The Seiberg-Witten curves for this class of theories were found in \cite{NP, Fucito:2012xc}, using a fairly involved analysis. There is a simple way to obtain the Seiberg-Witten curve: it can be derived from another curve, called the ${\cal S}$-curve in \cite{AHS}.
 The ${\cal S}$-curve is the Seiberg-Witten curve at the point where Higgs and the Coulomb branches meet. 
  
We know the Seiberg-Witten curve if we know the Higgs field $\varphi(x)$ solving \eqref{Hitch}, or equivalently \eqref{Bogomol}. On the Higgs branch, when all the monopoles become Dirac monopoles, solving the Bogomolny equations in \eqref{Bogomol} explicitly is easy,  and the ${\cal S}$ curve follows. 
The effect on $\varphi$ of adding a Dirac monopole of charge $\omega_i^{\vee}$, at a point $x=R' m_i$ on ${\cal C}$,  is to shift:
\beq\label{addone}
e^{R'\varphi} \rightarrow e^{R'\varphi}  \cdot (1-e^{x-R'm_i})^{-w_i^{\vee}}.
\eeq
This follows \cite{NP} by solving Bogomolny equations for $\phi$ on ${\cal C}\times S^1(R')$, and then dropping the dependence on positions on $S^1$. (In \cite{NP}, one had monopoles on a plane parameterized by $x$. We compactify $x$ into a cylinder ${\cal C}$ by adding infinitely many images to impose $x\sim x+2\pi i$. This amounts to replacing $x$ by $e^x$.) In the absence of Dirac monopoles, the $\varphi$ would have been constant, given by\footnote{Recall that, in IIB variables, $\varphi_a = (e_a, \varphi) = {\hat R}' m_s^2 \int_{S_a} (m_s^2 \omega_I/g_s + i B_{RR}) = \tau_a/R'$.} the vacuum value $R'\varphi = \tau$. (As vectors, we can simply identify the co-weights and weights, since ${\bf g}$ is simply laced. At times, we will want to keep the distinction in the labeling in order to recall that the Higgs field $e^{{\varphi_{\cal S}(z)}}$ lives in the maximal torus of ${\bf g}$.)
Thus, the Higgs field $\varphi = \varphi_{\cal S}(x)$ solving the Hitchin equation at the point where the Higgs and the Coulomb branches meet is
\beq\label{Higgs}
e^{R'\varphi_{\cal S}(x)} = e^{\tau} \prod_{\omega^V_i \in {\cal W}_{{\cal S}} }\;(1-e^{x-R'm_i})^{-\omega^\vee_{i}}.
\eeq
The ${\cal S}$ curve in representation ${Q}$ is simply obtained by specializing $\varphi=\varphi_{\cal S}(x)$ in \eqref{5dSWcurve},\eqref{4dSWcurve}. This amounts to finding the eigenvalues of $\varphi_{\cal S}$ in representation $Q$ and computing the determinant by taking the product. 
%

 From the ${\cal S}$-curve, we can recover the Seiberg-Witten curve at a generic point on the Coulomb branch by turning on  generic normalizable moduli. This was discovered and explained in \cite{DV, DV2, DVp, CKV, Cachazo:2001sg}.  
 
\subsubsection{Defects in $(2,0)$ CFT and Little String}
\label{subsubsec:defectlimit}

In the limit in which $(2,0)$ little string becomes $(2,0)$ CFT, we expect to get an ordinary Hitchin system on ${\cal C}$, with defects. The Higgs field $\varphi_{\cal S}$ corresponding to it is obtained from the solution in \eqref{Higgs} by taking a limit described in section \ref{subsubsec:CFTl}. We take $m_s$ to infinity, keeping $\varphi$, as well as  ${\hat R}$ and ${\hat R'}$, fixed in the limit.\footnote{The scalar $\varphi(x)$ equals, up to a factor of ${\hat R}'$, the modulus of the $(2,0)$ theory, $\varphi_{2,0}= m_s^2 \int_{S_a}(m_s^2 \omega_I + i B_{RR}) $.}  This means that we need to take $\tau$ to zero in the limit, since it equals to $R'$ times the value of $\varphi$ in the absence of monopoles. We will denote by $\alpha_0$ the vacuum value of $\varphi$,  $\alpha_0 = \tau /R'.$ A more subtle aspect of the limit is that we need to collide the fivebranes, by taking $R'$ to zero and keeping $\Delta x/R'$, their relative distances on ${\cal C}$, fixed. We can write the position $x_i$ of a D5 brane from $\omega_i$ as 
\beq\label{posi}
e^{x_{i}} = z_{\cal P} \;e^{R' \alpha_{i, {\cal P}}}, 
\eeq
and keep $z_{\cal P}$ and $\alpha$ fixed as we take $R'$ to zero.  Not all the D5 branes need to coalesce at the same point on ${\cal C}$; instead, we may group them into subsets of weights in ${\cal W}_{\cal P}$ that separately sum to zero 
$$\sum_{\omega_{i} \in {\cal W}_{\cal P} } \omega_i=0.$$
Because the sum of the $w_i$'s in ${\cal W}_{\cal P}$ vanishes, there are no redundant parameters in \eqref{posi}; shifting all $\alpha_i$'s by the same amount does not affect $\varphi$. It is easy to see that, in the limit where $m_s$ goes to infinity, we get:
$$
\varphi_{\cal S}(x) =  \alpha_0 +  \sum_{{\cal P}} \sum_{\omega_i \in {\cal W}_{{\cal P}}} {\; \alpha_{i,{\cal P}} \, \omega_i^{\vee}\over z_{\cal P} e^{-x} - 1}.
$$
It is more convenient to rewrite this in terms of a new variable $z=e^x$. Since $\varphi(x) dx$ is a one form on the Riemann surface, it the change of variables transforms it to $\varphi(z) =\varphi dx/dz = \varphi(z)/z$, and consequently,
\beq\label{Higgs}
\varphi_{\cal S}(z) = {\alpha_0\over z} +  \sum_{{\cal P}} \sum_{\omega_i \in {\cal W}_{{\cal P}} }\;{ \alpha_{i,{\cal P}} \, \omega_i^{\vee}\over z_{\cal P} - z}.
\eeq
This tells us that, in the $(2,0)$ CFT, we have poles on ${\cal C}$ at $z= z_{\cal P}$, with residues 
$$\alpha_{\cal P} =  \sum_{\omega_i \in {\cal W}_{{\cal P}}} \alpha_{i,{\cal P}} \, \omega_i^{\vee}.$$
%
This is the expected behavior of the Higgs field $\varphi$ near the defects in the $(2,0)$ theory on ${\cal C}$ \cite{G2}. As we will discuss in section \ref{sec:Examples}, when we take ${\cal W}_{\cal P}$ to correspond to a set of $n+1$ weights $\omega_i$ that span the weight lattice, the residue $\alpha_{\cal P}$ is generic, which leads to a full puncture at $z=z_{\cal P}$. It would be clearly important to connect the description of defects in the $(2,0)$ theory which we derived here, to the a-priori different description of defects in \cite{Tachikawa:2009rb, Chacaltana:2012zy}.

\subsection{Coulomb Branch for Finite and Infinite $m_s$}

At generic points on the Coulomb branch, $\varphi(x)$ is no longer diagonalizable. The Seiberg-Witten curve is no longer the same as the ${\cal S}$-curve, only their asymptotic behavior, near the punctures on ${\cal C}$, is the same. This is the case since Coulomb branch moduli are normalizable deformations of the Seiberg-Witten curve. The dimension of the Coulomb branch of little string theory is $ \sum_{a=1}^n (d_a-1)$, where $d_a$ are the ranks of the gauge groups in \eqref{5dgaugegroup}. This takes into the account that, while the Coulomb moduli coming from the $U(1)$ centers of the 5d gauge groups are not normalizable -- they affect residues of $\varphi$ at the puncture at $z=\infty$. 

The dimension of the Coulomb branch of the $(2,0)$ CFT compactified on ${\cal C}$ is generically smaller -- in the $m_s$ to zero limit, some deformations that were distinct at finite $m_s$ become indistinguishable. One way to count the moduli in the limit is from the perspective of IIB on $X\times {\cal C}$. In the $m_s$ to infinity limit, when the moduli space becomes $({\mathbb R})^5/W$, we can use an $R$ symmetry rotation to reinterpret  $\varphi_a =  \int_{S_a} m_s^4\omega_I/g_s+i m_s^2 B_{RR}$ as $\varphi_a =  \int_{S_a} m_s^4(\omega_J+i \omega_K)/g_s$. This does not change the theory in the limit, but the latter has a different geometric interpretation: it is a complex structure deformation of $X\times {\cal C}$, which turns it into a Calabi-Yau threefold $Y$, since $\varphi(x)$ varies over ${\cal C}$. This lets one use techniques of complex geometry to count the number of Coulomb moduli as the number of normalizable complex structure deformations of $Y$. For the resulting class of Calabi-Yau manifolds, the counting was done in \cite{CKV, Cachazo:2001sg}. For example, if one assumes that ${\cal C}$ starts out as a sphere with $k$ full punctures, where the residues $\alpha_{\cal P}$ of $\varphi(z)$ are all generic, the dimension of the Coulomb branch is $(k-2)h_+({\bf g})-n$, the where $h_{+}({\bf g})$ is the number of positive roots of ${\bf g}$. This is not surprising, as we are embedding the $(2,0)$ CFT into a theory with a lot more degrees of freedom.

\section{ADE  Little String and D3 Branes}

On the Higgs branch of little string theory, the bulk theory is abelianized, the D5 branes are all non-compact. At the same time, there is a new class of brane that plays an important role: these are D3 branes which are at points on ${\cal C}$ and which wrap compact 2-cycles in $X$. (The D3 branes wrapping non-compact 2-cycles are also important; they are codimension 4 defects of the little string, studied in \cite{AHC}.) The D3 branes survive the little string limit, for the same reason D5 branes in section 2 did; their tensions remain finite. 


\subsection{D3 Brane Gauge Theory}
\label{subsec:3dgauge}

String theory provides a derivation of the gauge theory on the D3 branes wrapping compact 2-cycles in $X$ in presence of non-compact D5 branes on cycles ${\cal W}_{\cal S}$. Let the homology class of the D3 branes in $H_2(X, {\mathbb Z})=\Lambda$ be
\beq\label{D3}
[D] = \sum_{a=1}^n N_a \;  e_a\qquad \in \qquad \Lambda
\eeq
where $N_a$ are positive integers. In the absence of D5 branes on $X$, the theory on the D3 branes was derived by Douglas and Moore in the seminal paper \cite{DM}. 
The theory is an ADE quiver theory with ${\cal N}=4$ supersymmetry in 3d, with
$U(N_a)$
gauge group for the $a$-th node of the quiver, leading to 
\beq\label{3dgauge}
\prod_{a=1}^n U(N_a),
\eeq
and $I_{ab}$ bufundamental matter hypermultiplets for each pair ${(a,b)}$ of nodes of the Dynkin diagram. From the ${\cal N}=2$ perspective, each vector multiplet contains an adjoint chiral multiplet, and there is a cubic superpotential.

It remains to deduce the effect of the D5 branes. Recall that, on the Higgs branch, the D5 branes wrap a collection of cycles $\{S_i^*\}$, whose homology classes make up ${\cal W}_{\cal S}$. Quantizing D3-D5 strings, we get chiral or anti-chiral multiplet of ${\cal N}=2$ supersymmetry for each intersection point in $X$ between the compact 2-cycle $S_a$ wrapped by the D3 branes and $S_i^*$ wrapped by the D5 branes. This follows since there are 6 Dirichlet-Neumann directions for open strings with one boundary on D5 branes and one on D3 branes: two come from the ${\mathbb R}^4$ part; four more DN directions come from the fact that the branes wrap two-cycles in $X$, intersecting transversally. The matter fields preserve 4 supersymmetries, since the D5 branes break half of the supersymmetry of $X$. This requires knowing the details of the geometry of the cycles wrapped by the branes. A much simpler quantity to determine is the net number of anti-chiral minus the chiral multiplets transforming in fundamental representation of the gauge group \cite{Cachazo:2001sg}:
\beq
\label{intersection}
\#(S_a, S_i^*) = ( e_a, \omega_i )
\eeq
for which we only need to know the homology classes $[S_a]=e_a$ of the compact and the non-compact cycles $[S_i^*]=\omega_i$ (the anti-chiral multiplets are the CPT conjugates of the chiral ones). In the examples that are relevant for us, the geometry of the cycles $S_i^*$ is simple, and the index \eqref{intersection} turns out to detects the full chiral matter content of the theory. Finally, the fact that the theory on D3 branes is really a 3d ${\cal N}=2$ gauge theory on a circle of radius $R$ follows most easily from T-duality that maps D3 branes at points on ${\cal C} = {\mathbb R} \times S^1(\hat R)$ to D4 branes wrapping $S^1(R)$.  We chose the moduli of $X$ so that $\int_{S_a} m_s \omega_{J}/g_s >0$ in \eqref{mod} and set the gauge couplings of the theory to zero on the D3 branes. The parameters $\tau_a$ in \eqref{taua} are the real FI terms, which are complexified because the theory is a 3d gauge theory on a circle. The remaining moduli in \eqref{mod} are the complex FI terms of the D3 brane gauge theory, which we set to zero.

\subsection{D3 Branes are Vortices}
\label{subsec:3dgaugev}

The D3 branes realize vortices in the D5 brane gauge theory. Vortices are co-dimension two solutions of gauge theories on the Higgs branch, where the vortex charge is the magnetic flux in two directions traverse to the vortex. A generic collection of vortices in 5d ${\cal N}=1$ gauge theories are BPS if the 5d FI parameters are aligned. At each node, the triplet of FI terms transform as a vector under the $SU(2)_R$ symmetry rotations. The orientation of this vector determines the supersymmetry preserved by the vortex.
In our setting, the 5d FI parameters are the moduli of the little string in \eqref{FI}. The background we consider has $\int_{S_a} m_s^2 \omega_{J}/g_s >0$ as the only non-zero FI terms in \eqref{FI}. The vortices are in fact the supersymmetric vacua of the theory on the D3 branes. Due to non-zero 3d FI terms (recall that ${\rm Re}(\tau_a)>0$) in a supersymmetric vacuum, the chiral multiplets from the D3-D5 strings need to get expectation values. This describes D3 branes ending on the D5 branes. As is well known, this turns on magnetic flux on the D5 brane, transverse to the D3 branes \cite{HW}, consistent with the vortex interpretation.

The Higgs branch of the theory on the D3 branes, from the previous section, is the moduli space of vortices. We have thus derived, from string theory, the description of the moduli spaces of  vortices in a large class of ADE quiver ${\cal N}=2$ theories. As far as we are aware, the result is novel, except in some special cases. (In mathematical literature, the moduli space of vortices is called the moduli space of quasi-maps; see for example \cite{Nakajima}, where the quiver in figure \ref{fig:AnQuiver} appeared before, precisely for this purpose.) 

\subsection{Partition Function of D3 Branes}
\label{subsec:3dpart}
Consider now the partition function of the D3 brane theory in $\Omega$-background. From the perspective of the little string, the partition function corresponds to the theory in exactly the same background as in section \ref{subsec:5dpart}, except now the bulk theory is on the Higgs branch, and all the D5 branes are non-compact. In addition we have $N_a$ D3 branes wrapping the 2-cycle $S_a$ and the complex plane rotated with parameter $q$; we take the plane rotated by $t$ to be transverse to the branes. Of course, for the same reason the theory on the D5 branes wrapping 2-cycles in $X$ was a five dimensional ${\cal N}=1$ gauge theory, the theory on the D3 branes on 2-cycles is a three dimensional gauge theory with ${\cal N}=2$ supersymmetry, both on a circle of radius $S^1(R)$.

The partition function of the theory ${Z}_{{3d}}$ is
\beq\label{3dtrace}
Z_{3d} = {\rm Tr} (-1)^F \; { g},
\eeq
where as before, ${g} =q^{S_1 - S_R} t^{-S_2 + S_R}$ is a combination of $SO(2)$ rotations $S_{1,2}$ of the two copies of ${\mathbb C}$.  $S_R$ is the same R-symmetry rotation as in section \ref{subsec:5dpart}; we chose the background in section \ref{subsec:3dgauge} so that the symmetry is preserved. Before we add D5 branes, the theory has ${\cal N}=4$ supersymmetry. Then, the generators we call $S_R$ and $S_2$ are the generators of the $U(1)_H\times U(1)_V$ subgroup of the $SU(2)_H\times SU(2)_V$ R-symmetries that act on the Higgs and the Coulomb branches of the theory, respectively. This identification comes from the fact that $S_2$ acts by phase rotation of the ${\cal N}=2$ chiral adjoint multiplet which sits inside the ${\cal N}=4$ vector multiplet, and does not act on the hypermultiplets. The bifundamental hypermultiplet transforms as a doublet of $SU(2)_H$. From the perspective of the 3d gauge theory, if we take $S_1$ to rotate the D3 brane world-volume, and $S_2$ to rotate the space transverse to brane, then both $S_2$ and $S_R$ are R-symmetries. Since the theory on the D3 branes has 3d ${\cal N}=2$ supersymmetry that has at most a $U(1)_R$ symmetry, the difference $S_R-S_2$ is in fact a global symmetry. 

The partition function can be computed as the integral over the Coulomb branch:
\beq\label{3dpart}
{Z}_{{3d}}=\int dx \; I_{{3d}}(x).
\eeq
The $x$'s are the Coulomb moduli of  the 3d gauge theory, the positions on ${\cal C}$ of the D3 branes. 
The integrand $I_{{3d}}(x)$ is the contribution to the index \eqref{3dtrace} from one loop integrating out of massive charged matter fields (detailed study of partition functions of this type is in \cite{BDP, Shadchin:2006yz, Hori:2013ika, Honda:2013uca, Yoshida:2014ssa}), together with classical terms. It can be read off from the quiver, by including contributions of all gauge and matter fields
\beq\label{basic}
I_{3d}(x) = r_{3d} \; \prod_{a=1}^n I^{3d}_a(x_a)\cdot I^{3d}_{a, V}(x_a, f) \cdot \prod_{a<b } I^{3d}_{ab}(x_a, x_b)
\eeq
The contribution of the ${\cal N}=4$ $U(N_a)$ vector multiplets is
\beq\label{basic1}
I^{3d}_a(x_a)= e^{\sum_{I=1}^{N_a} \tau_a x_{a,I}}\; \prod_{1\leq I \neq J \leq N_{(a)}} { \varphi_q(  e^{x^{(a)}_I-x^{(a)}_J})\over \varphi_q(   t \,e^{x^{(a)}_I-x^{(a)}_J})},
\eeq
where the numerator comes from $W$-bosons, the denominator from the adjoint chiral multiplet within the vector multiplet.
The bifundamental hypermultiplets, corresponding to a pair of nodes $a$ and $b$, give:
\beq\label{basic2}
 I^{3d}_{ab}(x_a, x_b)= \prod_{1\leq I \leq N_{(a)}}   \prod_{1\leq J\leq N_{(b)}} \Bigl(  {\varphi_q(  v\, t \, e^{x^{(a)}_I-x^{(b)}_J})\over \varphi_q( v \,e^{x^{(a)}_I-x^{(b)}_J})}\Bigr)^{I_{ab}},
\eeq
The contribution is non-trivial only for the pairs of nodes connected by a link in the Dynkin diagram.
For every $\omega_i \in {\cal W}_{\cal S}$, we get a collection of chiral multiplets and anti-chiral multiplets. A chiral multiplet in fundamental representation of the gauge group on the $a$-th node, with $S_R$ R-charge $-r/2$ contributes  $\prod_{1\leq I \leq N_a}\Bigl( \varphi_q( v^{r} f/ e^{x^{(a)}_I}) \Bigl)^{-1}$, 
while an anti-chiral multiplet in fundamental representation contributes $\prod_{1\leq I \leq N_a}\Bigl( \varphi_q( v^{r} f/  e^{x^{(a)}_I}) \Bigl)$, where $f$ is the flavor fugacity and $v=\sqrt{q/t}$. The function $\varphi_q(z)$ is as in section \ref{subsec:5dpart}.
For each node, we get a contribution of the form
$$
I^{3d}_{a, V}(x_a, f)=   \prod_{w_i \in {\cal W}_{\cal S}} I^{3d}_{a, \omega_i}(x_a, f_i)
$$
where
\beq\label{basic3}
I^{3d}_{a, \omega_i}(x_a, f_i)
\eeq
captures the contributions of all the chiral and anti-chiral matter fields coming from strings stretching between the D5 brane wrapping $[S_i^*]=\omega_i$ and the D3 brane on $e_a=[S_a]$. This depends on  $f_i = \exp{(R m_i)}$, encoding the position $x_i=Rm_i$ of the D5 brane on ${\cal C}$.  To write down the explicit formula in \eqref{basic3}, one in general needs to know the spectrum, and not just the index \eqref{intersection}. The ${\cal N}=4$ matter contribution in \eqref{basic2}, by contrast,  is fixed by supersymmetry. For the theories in Fig.1 we will give the explicit formulas in section \ref{sec:Examples}. The integral runs over the Coulomb branch moduli for each of the $n$ gauge group factors in \eqref{3dgauge}:
\beq
"\int dx\;" =
{1\over |W_{G_{3d}}|}\;\;\prod_{a=1}^n\; \int d^{N_a}x_{(a)}.
\eeq
We still need to specify the integration cycle in \eqref{3dpart}. The integration cycles, in turn, correspond to vacua of the 3d gauge theory, see for example \cite{BDP}. We will postpone discussing the contours until section 5, when we will need them.

\section{$ADE$ Toda and its $q$-deformation}
\label{sec:Toda}

The D3 branes on $X$ have a close relation to a 2d conformal field theory. The partition function of the gauge theory on D3 branes in \eqref{3dpart} turns out to be equal to a certain canonical {"$q$-deformation"} of the ADE Toda CFT conformal block on ${\cal C}$ with ${\cal W}_{q,t}({\bf g})$  vertex algebra symmetry, found by Frenkel and Reshetikhin in \cite{FR1}. The relation between them is manifest, as soon as one recalls basics of the Toda CFT, and the construction in \cite{FR1}.
Moreover, taking the $m_S$ to infinity limit that brings $(2,0)$ little string back to $(2,0)$ CFT, the $q$-deformed Toda CFT reduces to the ordinary Toda CFT, and  ${\cal W}_{q,t}({\bf g})$ to the ${\cal W}({\bf g})$ algebra. However, just as was the case for the D5 branes, the limit does not correspond to a partition function of any gauge theory with a Lagrangian -- the relation between the two is simple only within the little string theory.

\subsection{Free Field Toda CFT}

The ADE Toda field theory can be written in terms of $n={\rm rk}({\bf g})$ free bosons in two dimensions with a background charge contribution and the Toda potential that couples them:
\beq\label{Todaaction}S_{Toda} =  \int dz d{\bar z} \;\sqrt g \; g^{z{\bar z}}[(\partial_z \varphi,  \partial_{\bar z} \varphi)+  (\rho, \varphi)\,Q R + \sum_{a=1}^n e^{(e_a,\varphi)/b} ].
\eeq
The field $\varphi$ is a vector in the $n$-dimensional (co-)weight space, $(,)$ is the Killing form on the Cartan subalgebra of $\bf g$,  $\rho$ is the Weyl vector, and $Q=b+ 1/b$.  As before, $e_a$ label the simple positive roots.
The Toda CFT has an extended conformal symmetry, a ${\cal W}({{\bf g}})$ algebra symmetry. (For a review of ${\cal W}$ algebras see \cite{Bouwknegt:1992wg}. The free field formalism for Toda CFT was discovered in \cite{Dotsenko:1984nm} and studied in the present context in \cite{DVt, Itoyama:2009sc,Mironov:2010zs,Morozov:2010cq, Maruyoshi:2014eja}). The primary vertex operators of the ${\cal W}({{\bf g}})$ algebra are
labeled by an $n$-dimensional vector of momenta $\alpha$, and given by:
\beq\label{primary}
V_{\alpha}(z) = e^{(\alpha, \varphi(z))}.
\eeq
The free field conformal blocks of the Toda CFT have a particularly simple form:
\beq\label{expect}
\langle V_{\alpha_1}(z_1) \ldots V_{\alpha_k}(z_k) \;\;\prod_{a=1}^{n} Q_{a}^{N_a} \rangle_{free}
\eeq
where screening charges
$$
Q_{(a)} = \oint dx \,S_{a}(x)
$$
are the integrals over the screening current operators $S_a(x)$, one for each simple root,
\beq\label{scc}
S_{(a)}(z) = e^{ (e_{a}, \phi(z))/b}.
\eeq
We will only give a rough sketch of the derivation of \eqref{expect} (see \cite{Fateev:2007ab} for details). Consider treating the Toda potential as a perturbation, expanding and bringing down powers of the Toda potential. Each term is now a computation of the correlation in a free field theory, with insertions of screening charge integrals coming from Toda potential. Momentum conservation picks out a single term, the one with the net vertex operator momentum plus the momenta of the screening charges, 
\beq\label{constraint}
\sum_{i=1}^k \alpha_i+ \sum_{a=1}^n N_a e_{a}/b = 2Q.
\eeq 
The last term comes from the background charge on a sphere, induced by the curvature term in \eqref{Todaaction}. Picking out the chiral half of the correlation, results in \eqref{expect}. A more precise derivation of the constraint results directly from the path integral, by integrating over the zero modes of the bosons \cite{Fateev:2007ab}. The constraint \eqref{constraint} says that one of the momenta, say the momentum $\alpha_{\infty}$ of the vertex operator at $z={\infty}$, is fixed in terms of momenta $\alpha_i$ of all the other vertex operators and numbers of screening charge integrals $N_a$.
Computing the correlators using Wick contractions, the conformal block has the form of an integral over the positions $x$ of $N_a$ screening current insertions
\beq\label{conf1}
Z_{Toda} = \int dx \;I_{Toda}(x)
\eeq
where the integrand $I_{Toda}$, coming form \eqref{expect} is a product over two-point function of the screening currents with themselves, running over all pairs of nodes of the Dynkin diagram, and two-point functions of screening currents and vertex operators:
\beq\label{cbasic}
I_{Toda}(x) = \prod_{a=1}^n I^{Toda}_a(x_a)\cdot I_{a, V}(x_a, z) \cdot \prod_{a<b } I^{Toda}_{ab}(x_a, x_b)
\eeq
The two-point functions of screening currents at a fixed node contribute 
\beq\label{cbasic1}
I^{Toda}_{a} = \prod_{1\leq I \neq J \leq N_{(a)}}  \langle S_a(x^{(a)}_I), S_a(x^{(a)}_J) \rangle_{free}.
\eeq
It will become apparent momentarily that, in the $q$-deformed theory, these directly correspond to ${\cal N}=4$ $U(N_a)$ vector multiplet contributions coming from $U(N_a)$ gauge theory at $a$-th node. The two-point functions of screening currents between distinct nodes $a$ and $b$ contribute
\beq\label{cbasic2}
I^{Toda}_{ab} = \prod_{1\leq I\leq  N_{(a)}}\prod_{1\leq J\leq N_{(b)}}  \langle S_a(x^{(a)}_I), S_b(x^{(b)}_J) \rangle_{free}.
\eeq
These correspond to bifundamental hypermultiplet contributions. Finally, the two-point functions of screening currents at a given node with all the vertex operators,
\beq\label{cbasic3}
 I^{Toda}_{a, V} = \prod_{i=1}^k \prod_{1\leq I \leq N_{(a)}}\   \langle S_a(x^{(a)}_I),  V_{\alpha_i}(z_i) \rangle_{free},
\eeq
will correspond to chiral matter contributions. 
The two-point functions  are those of ordinary free bosons, so they are simply equal\footnote{We have been cavalier throughout with the momentum zero modes. It is a straightforward exercise to restore these.} 
 to:
\beq\label{sct}
 \langle S_a(x), S_b(x') \rangle_{free} = (x-x')^{b^2C_{ab} }
\eeq
\beq\label{vt}
\langle S_a(x), V_{\alpha} (z) \rangle_{free} = (x-z)^{(\alpha, e_a)}.
\eeq
The relation to a gauge theory arises only after $q$-deformation. 
The structure of the partition function as an integral in \eqref{conf1}, \eqref{cbasic}, is reminiscent of the 3d gauge theory partition function in \eqref{3dpart}, \eqref{basic}. After we $q$-deform the Toda CFT, they become the same.  
\vskip 0.5 cm
{\it Contours and Fusion Multiplicities}
\vskip 0.5cm
To fully specify the conformal block, we need to make a choice of contour in \eqref{conf1}. Conformal blocks can be obtained as solutions to differential equations of hypergeometric type, and \eqref{conf1} can be viewed as providing an integral solution to the equation, see for example \cite{Fateev:2005gs,Fateev:2007ab}. Generically, there is a finite dimensional space of solutions to such equations, and choosing a contour picks out a specific one. We will compute the dimension of the space of conformal blocks of ${\cal W}({\bf g})$ algebra, following \cite{Fateev:2007ab, Wyllard:2009hg, Fateev:2010za}, where the calculation was done for ${\bf g} ={ A_n}$. The ${\cal W}({\bf g})$ algebra is a vertex operator algebra with $n$ generators ${\cal W}^{(s_a)}$, labeled by their spins $s_a$. It has Virasoro algebra as a subalgebra, generated by the spin 2 generator, the stress tensor. The vertex operators $V_{\alpha}(z)$ in \eqref{primary} are primaries of the entire ${\cal W}$ algebra, not just of its Virasoro subalgebra. The Virasoro symmetry suffices to determine the correlation functions of all the Virasoro descendants in terms of those of the Virasoro primaries. But, since ${\cal W}$ algebra is bigger than the Virasoro algebra, the set of Virasoro primaries is larger than the set of ${\cal W}$ algebra primaries -- it includes not only $V_{\alpha}(z)$, but also their ${\cal W}$ algebra descendants. One can use the ${\cal W}$ algebra and the Ward identities to reduce the space of the descendant insertions.
For the $k$-point function on a sphere, involving $k$ primary operators of generic momenta, the number of ${\cal W}$ algebra generators that cannot be removed is $(k-2)h_+ - n$ \cite{Fateev:2010za}, where $h_+ = h_+({\bf g})$ is the number of positive roots of ${\bf g}$. This is the additional data one needs to specify besides the $k$ external momenta $\alpha_i$ to specify the block in \eqref{conf1}. The calculation is  straightforward: The spins $s_a$ of the ${\cal W}$ algebra generators are determined by the group theory \cite{Bouwknegt:1992wg}. They are given by the {\it exponents} of ${\bf g}$ augmented by $1$. For a spin $s$ generator $W^{(s)}(z)$, the ${\cal W}$-algebra can be used to remove all but the $s-1$ of its modes $W^{(s)}_{-s+m}$, where $m$ runs from $1$ to $s-1$. There is a global Ward identity on a genus zero Riemann surface that imposes $2s-1$ linear relations between the insertions of $W^{(s)}$'s at different points, further reducing the multiplicity to $k(s-1)-(2s-1)$ \cite{Bouwknegt:1992wg, Fateev:2007ab}. For ADE Lie algebras, the exponents are easily seen to satisfy 
\beq\label{exponentsum}
\sum_{a=1}^n (s_a-1) = h_+,
\eeq
leading to the result we quoted.

\subsection{Toda CFT, $q$-deformation and ${\cal W}_{q,t}({\bf g})$-algebra}

The ${\cal W}({\bf g})$-algebra can be defined as a complete set of currents that commute with the screening charges. In \cite{FR1}, one constructed a deformation of both the algebra and the conformal blocks in free field formulation, based on deforming the screening currents. The screening current $S_a(x)$ operators are deformed 
so that 

\beq\label{sctda}
 \langle S_a(x), S_a(x') \rangle_{free} = { \varphi_q(  e^{x-x'})\over \varphi_q(   t \,e^{x -x'})}  { \varphi_q(  e^{x'-x})\over \varphi_q(   t \,e^{x' -x})},
\eeq
and, for $a\neq b$
\beq\label{sctdab}
 \langle S_a(x), S_b(x') \rangle_{free} = \Bigl({ \varphi_q( tv e^{x-x'})\over \varphi_q(   v \,e^{x -x'})}\Bigr)^{I_{ab}} 
 \eeq
where $v = (q/t)^{1\over 2}$.
The explicit formulas the $q$-deformed screening charges $S$ are in \cite{FR1}, and we review them in the appendix A. The ${\cal W}_{q,t}({\bf g})$ algebra is defined in \cite{FR1} as a set of all operators commuting with the deformed screening charges  (together with a set of screening charges with $q$ and $t$ exchanged). After the $q$-deformation, the conformal block \eqref{cbasic} becomes manifestly equal to the partition function of the D3 brane gauge theory: the screening charge contributions  in \eqref{cbasic1}, \eqref{cbasic2} to the conformal block are the contributions of ${\cal N}=4$ vector and hypermultiplets in \eqref{basic1} and \eqref{basic2} to the D3 brane partition function, where the number $N_a$ of D3 branes on $a$-th node maps to the number of screening charge insertions.

The  primary vertex operators in the q-deformed ${\cal W}_{q,t}({\bf g})$ algebra are deformations of \eqref{scc}. A construction of $q$-deformed primary vertex operators, with generic momenta, is given in the appendix A. They have the form
$$
:\prod_{i=1}^{n+1} V_{{\omega}_i}(x_i): \rightarrow V_{\alpha}(z)
$$
where $\omega_i$ are a collection of $n+1$ weights of ${\bf g}$, suitably chosen. See appendix A and sectio \ref{sec:Examples} for explicit expressions. Each of the $V_{\omega_i}(x)$'s 
has two point functions  with the screening operators that equal \eqref{basic3},
\beq\label{sv}
 \langle S_a(x) V_{\omega_i}(f) \rangle_{free}.
 \eeq
Explicitly, $V_{\omega_i}(f)$ is a normal ordered product of fundamental vertex operators of the form $W^{\pm 1}_{a}(f v^{r})$ whose two point functions with the screening charges are $\varphi_q^{\pm 1}(f v^r/x)$, equal to the contributions to \eqref{basic3} of either chiral or anti-chiral multiplets of $R$-charge $r$. To go back to the undeformed theory, we can let
$q= \exp(R \epsilon_1), t=\exp(-R \epsilon_2)$, and take the $R$ to zero limit.  In the limit, $q$ and $t$ tend to $1$, where we take \eqref{posi}
\beq\label{cftposm}
e^{x_{i}} = z_{\cal P} \;q^{ \alpha_{i, {\cal P}}}, 
\eeq
\eqref{sv} above becomes $(1- e^x/z_{\cal P})^{(e_a, \alpha_{\cal P})} $ where $\alpha_{\cal P} =\sum_{w_i \in {\cal W}_{\cal P}} x_i \; w_i $. This is the two-point function of the vertex operator with the screening charge in Toda CFT, given in \eqref{sct}. In principle, one can consider insertions of any collection ${\cal W}_{\cal S}$ of $V_{\omega_i}(x)$'s with $\sum_{\omega_i \in {\cal W}_{\cal S}} \omega_i=0$. The CFT limit of this depends on the collection ${\cal W}_{\cal S}$, and how insertion scales in the $R$ to zero limit: one can get any collection of primary vertex operators with either arbitrary or (partially) degenerate momenta.

 After the $q$-deformation, the ${\cal W}({\bf g})$ algebra is no longer a symmetry, since ${\cal W}({\bf g})$ is not a subalgebra of ${\cal W}_{q,t}({\bf g})$, so the argument does not apply. Correspondingly, instead of hypergeometric equations, the conformal blocks now satisfy $q$-difference equations. The number of solutions of these can be larger, as some linearly independent solutions to $q$-hypergeometric equations can become linearly dependent in the $q\rightarrow 1$ limit. We will see that this is indeed what happens for ${\bf g}\neq A_n$.

\section{Triality}
\label{sec:Triality}

We will show that the partition function of $(2,0)$ little string on ${\cal C}$ with codimension two defects equals the $q$-deformation of the Toda CFT conformal block on ${\cal C}$, with vertex operators determined by positions and types of defects. 
%
%
Since we have shown, in section 6, that the $q$-deformed CFT correlator \eqref{conf1} equals the 3d partition function in \eqref{basic}, we only need to show equality of partition functions of the 3d gauge theory on D3 branes and the partition function of the $(2,0)$ theory on ${\cal C}$. 

The relation between the partition function of the $A_1$ 6d $( 2,0)$ CFT and the 2d Liouville CFT was conjectured by Alday, Gaiotto and Tachikawa  in \cite{AGT}. That conjecture was proven in \cite{Fateev:2009aw, Alba:2010qc}. Generalization of the correspondence for other groups were studied by many papers, see for example \cite{Wyllard:2009hg, Td,Keller:2011ek} and \cite{Teschner:2014oja} for a recent collection of reviews. For pure gauge theories of ADE type, the relation between the gauge theory partition function and the norm of the Whittaker vector of the ${\cal W}$-algebra was proven recently in \cite{1406.2381} (see \cite{Taki:2014fva} for the $q$-deformed $A_n$ version). 
An obstacle to extending the correspondence to groups other than $A_1$ is that compactification of $(2,0)$ 6d CFT on a Riemann surface does not in general lead to a theory with a Lagrangian description: without it, the partition function of the theory is not computable either, so there is nothing to compare to the Toda conformal block. Another obstacle is that, for ${\bf g}\neq A_1$, the general Toda conformal blocks are known only if they admit free field representation; after $q$-deformation, this is true even in the Virasoro case. 


The little string perspective of the present paper is crucial to establish the precise statement of the correspondence for arbitrary ADE groups and general blocks on a sphere, and leads to a proof of it for conformal blocks admitting free field representation, generalizing \cite{AHS} for $A_n$ (and  \cite{ACHS} for $A_1$). The correspondence between the $(2,0)$ theory and Toda theory in the conformal limit follows by taking the $m_s$ to infinity limit.

\subsection{Gauge/Vortex Duality}
The relationship between the 5d ${\cal N}=1$ gauge theory on D5 branes in section 3, and the 3d  ${\cal N}=2$ gauge theory on D3 branes in section 4 is gauge/vortex duality. The duality comes from two different, yet equivalent descriptions of vortices in the theory: one from the perspective of  5d theory with vortices, and the other from the perspective of the 3d theory on the vortex. 
The fact that the  theory on vortices captures aspects of dynamics of the "parent" gauge theory was noticed early on in \cite{Dorey} at the level of BPS spectra.  Turning on $\Omega$-background {\it transverse} to the vortex, the correspondence becomes more extensive \cite{DH1,DH2}: it is a gauge/vortex duality \cite{ACHS, AHS, ASv}.

The 2d $\Omega$-background transverse to the vortex (with parameter $\epsilon_2$) is necessary. It ensures that the super-Poincare symmetries preserved by the 5d and the 3d theories are the same, since the $\Omega$-background is a form of compactification \cite{NS, NS1, NS2} and breaks half the supersymmetry: after turning it on, both theories are 3d ${\cal N}=2$ theories on a circle. The duality should hold at the level of F-type terms -- the Kahler potentials are not protected, and we don't claim to specify them. The duality is the little string analogue \cite{ASv} of large $N$ dualities in topological string \cite{DV, DV2, DVp, CKV}. The D3 brane gauge theory lives in the Higgs phase of the bulk theory.  From the bulk perspective, the theory starts out on the Higgs branch, but ends up pushed onto the Coulomb branch due to the vortex flux. In the Higgs phase, the Coulomb moduli are frozen to points where the hypermultiplets can get expectation values. Turning on $N$ units of vortex flux in a $U(1)$ gauge group shifts the corresponding Coulomb modulus $a$ to $a+N \epsilon_2$, where $\epsilon_2$ is the parameter of the $\Omega$ in background rotating the complex plane transverse to the vortex. This is a consequence \cite{ACHS, AHS, ASv}
of how $\Omega$-background deforms the Lagrangian of the 5d theory \cite{NO, NW}.
Once we have a pair of dual theories, one expects that their partition functions in the full $\Omega$-background, depending on $\epsilon_{1,2}$, agree as well. This was shown in detail in \cite{ACHS, AHS} for the $A_1$ and the $A_n$ theories. The generalization to the ADE case works in an analogous way, so we will be brief.
\subsection{Equality of Partition Functions}
\label{subsec:Equality}

In the partition function of D3 branes in  \eqref{3dpart}, we choose an integration cycle, corresponding to choosing a vacuum of the theory.  Evaluating the contour integral, we pick up all the poles from \eqref{basic} within it. The poles turn out to be labeled by touples of 2d partitions $\{R\}$, with one row per integration variable. 
This allows us to express the contour integral in \eqref{3dpart} as a sum over 2d partitions
 \beq\label{3dparta}
  Z_{3d} = \oint dx\, I_{3d}(x) = \sum_{\{R\}} {\rm res}_{\{ R\}} I_{3d}(x).
 \eeq
 The residue at the pole labeled by $\{R\}$, normalized by the residue of the pole at $\{\varnothing\}$, corresponding to trivial 2d partition, is simply equal to the ratio 
 \beq\label{rat}
{\rm res}_{\{ R\}} \, I_{3d}(x)/ {\rm res}_{\{\varnothing\}}I_{3d}(x) =I_{3d}(x_{\{R\}})/I_{3d}(x_{\{\varnothing \}})
\eeq
of the integrands at the two points, which is finite and simple to compute.

From the bulk perspective, the choice of integration cycle in \eqref{3dpart} corresponds to picking the values of the Coulomb moduli where \eqref{5dp} gets evaluated, or more precisely, choosing the vortex fluxes which determine them. Evaluating the 5d partition function at these values \eqref{Coulomb}, one finds that it simplifies: only the partitions with finite numbers of rows, determined by the fluxes, contribute. 
The equality of the bulk partition function in \eqref{5dp} 
\beq\label{5dpa}
Z_{5d} = r_{5d} \sum_{\{R\}}\; I_{5d,\{R\}}(e)
\eeq
and the D3 brane partition function follows by noticing that term by term, contributions of a tuple $\{R\}$ to \eqref{5dp} equals to the 
residue in \eqref{3dparta}:
\beq\label{central}
I_{5d, \{R\}} = I_{3d}(x_{\{R\}})/I_{3d}(x_{\{\varnothing \}}).
\eeq
From the Toda CFT point of view, this corresponds to specifying the conformal block;  for $g\neq A_1$, there is a finite dimensional space of conformal blocks to choose from, even for the 3-point function, see section  \ref{sec:Toda}.

{\vskip 0.5 cm}
{\it Bulk Partition Function}
{\vskip 0.5 cm}

The bulk partition function \eqref{5dp}  is written in terms of the basic building block, the Nekrasov function in \eqref{nekrasovN}. We can rewrite the function as
\begin{align*}
{\cal N}_{R_1R_2}\Big( \frac{e_1}{e_2} \Big) = \ 
&
\prod\limits_{i = 1}^{N_1} \prod\limits_{j = 1}^{N_2} 
\dfrac{\varphi_q\big( \frac{e_1}{e_2} q^{R_{1,i}-R_{2,j}} t^{j-i + 1} \big)}{\varphi_q\big( \frac{e_1}{e_2} q^{R_{1,i}-R_{2,j} }t^{j-i} \big)} \ \dfrac{\varphi_q\big( \frac{e_1}{e_2} t^{j-i} \big)}{\varphi_q\big( \frac{e_1}{e_2} t^{j-i + 1} \big)} \\
& N_{R_1, \varnothing}\Big( t^{N_2} \frac{e_1}{e_2} \Big) N_{\varnothing, R_2}\Big( t^{- N_1} \frac{e_1}{e_2} \Big)
\end{align*}
assuming that the partitions $R_1$ and $R_2$ have no more that $N_1$ and $N_2$ rows, respectively. Let's suppose that a partition $R_{(a), I}$ in \eqref{5dp} has no more than $N_{(a), I}$ rows. Then, using the property just quoted, we can re-write the contributions of vector and hypermultiplets to \eqref{5dp}  as follows:
$$\prod_{a=1}^n z_{V_a, {\vec R}_{(a)} }=  \prod_{a =1}^n {I^{3d}_{a}(x_{\vec R_{(a)}})\over I^{3d}_{a}\big(x_{\vec\varnothing_{(a)}}\big)} \cdot V_{\rm vect},
$$
where $I_{3d, V_a}$ is the contribution of the 3d vector multiplet corresponding to a $U(N_{(a)})$ gauge group in \eqref{basic1} evaluated at positions 
\beq\label{polepositions}\{ e^{x_{\vec R_{(a)}}}\}  = \{e_{(a),I} \,q^{R_{(a),I}}\,t^{\rho}  v^{\#_a}  \},
\eeq
determined by the lengths of the rows of the partitions $\{R_{a,I}\}_{I=1}^{d_a}$.  Here, $ v^{\#_a} $ is a power of $v$ that depends on the position of the node in the Dynkin diagram: it is equal to $1$ for $a=1$, and increases by one with every link.
We labeled the remaining factor $V_{\rm vect}$. Similarly, the contributions of 5d bifundamentals can be written as
\begin{align*}
\prod^n_{a,b=1}z_{H^{a,b}, {\vec R}^{(a)},  {\vec R}^{(b)}} &= \prod_{a,b =1}^n  \Big[{ I^{3d}_{{ab}}\big(x_{\vec R_{(a)}},x_{\vec R_{(b)}} \big)\over I^{3d}_{ab}\big(x_{\vec\varnothing_{(a)}}, x_{\vec\varnothing_{(b)}}\big)} \Big]^{I_{ab}}\cdot \,V_{\rm bifund}
\end{align*}
where $I^{3d}_{ab}\big(x^{(a)} ,x^{(b)} \big)^{I_{a,b}}$ is the contribution of the 3d bifundamental hypermultiplet multiplet,  and $V_{\rm bifund}$ stands for all the remaining factors. 
We write the contributions of fundamentals as
$$\prod_{a=1}^n z_{H_a, {\vec R}^{(a)}}  =V_{\rm fund} 
$$
and $\prod_{a=1}^n z_{CS, {\vec R}^{(a)}} = V_{\rm CS}$,  for the 5d Chern-Simons contribution.
This lets us write \eqref{5dp}, the contribution of a the couple $\{R\}$ to the bulk partition function, as
$$
I_{5d,\{R\}}= \prod_{a =1}^n {I^{3d}_{a}(x_{\vec R_{(a)}})\over I^{3d}_{a}\big(x_{\vec\varnothing_{(a)}}\big)}\cdot \prod_{a,b =1}^n  \Big[{ I^{3d}_{{ab}}\big(x_{\vec R_{(a)}},x_{\vec R_{(b)}} \big)\over I^{3d}_{ab}\big(x_{\vec\varnothing_{(a)}}, x_{\vec\varnothing_{(b)}}\big)} \Big]^{I_{ab}}\cdot V_{\rm vect} V_{\rm bifund} V_{\rm fund} V_{\rm CS}
$$
This is merely a rewriting of \eqref{5dp}, in particular, the sum runs over arbitrary tuples $\{R\}$. 
However, several remarkable things happen if we set $e_{(a),I}$ equal to $f_{i} \,t^{N_{(a),I}},$ with a suitable proportionality constant:
\beq\label{Coulomb}
e_{(a),I} = f_{i} \;t^{N_{(a),I}}\, v^{\#_{a, i, I}}.
\eeq
Firstly, the product $V_{\rm vect} V_{\rm bifund} V_{\rm fund} V_{\rm CS} $
vanishes if partition $R_{(a),I}$ in the tuple has more than $N_{(a),I}$ rows. Secondly, there are many cancellations in the product, which simplifies to
\beq\label{longproducta}
V_{\rm vect} V_{\rm bifund} V_{\rm fund} V_{\rm CS} =\prod_{a=1}^n \Big[{ I^{3d}_{{a,V}}\big(x_{\vec R_{(a)}} \big)\over I^{3d}_{{a,V}}.
\big(x_{\vec \varnothing_{(a)}} \big)}\Big]
\eeq
The summands of the bulk partition function \eqref{5dp} turn into the D3 brane partition function, 
\beq\label{longcentral}
I_{5d,\{R\}}= \prod_{a =1}^n {I^{3d}_{a}(x_{\vec R_{(a)}})\over I^{3d}_{a}\big(x_{\vec\varnothing_{(a)}}\big)}\cdot \prod_{a,b =1}^n  \Big[{ I^{3d}_{{ab}}\big(x_{\vec R_{(a)}},x_{\vec R_{(b)}} \big)\over I^{3d}_{ab}\big(x_{\vec\varnothing_{(a)}}, x_{\vec\varnothing_{(b)}}\big)} \Big]^{I_{ab}}\cdot \prod_{a=1}^n \Big[{ I^{3d}_{{a,V}}\big(x_{\vec R_{(a)}} \big)\over I^{3d}_{{a,V}}
\big(x_{\vec \varnothing_{(a)}} \big)}\Big]
\eeq
evaluated at discrete set of points $x=x_{\vec R_{(a)}}$, determined from \eqref{Coulomb} and \eqref{polepositions}. The right hand side is the contribution of poles in evaluating the 3d partition function by contours. The family of points $x=x_{\vec R_{(a)}}$, indexed by 2d partitions of finite length, are the positions of the poles in the integral. This leads to the third remarkable fact:  The subtle $v$ factors are related to the $S_R$ R-symmetry charges of 3d chiral multiplets. These should be fixed by requiring superconformal invariance of the 3d theory in the IR. At the same time, they turn out to be determined by the ${\cal W}_{q,t}({\bf g})$ algebra symmetry in \cite{FR1}: the choice of the factors that equates the residues with the 5d partition function is exactly the same one as what is needed for the partition function to equal the $q$-deformed conformal block of the Toda CFT. The powers $v^{\#_{a, i, I}} $ in \eqref{Coulomb} can be read of directly from the vertex operators.


{\vskip 0.5 cm}
{\it 3d Partition Function}
{\vskip 0.5 cm}

The contours of integration in \eqref{3dparta} are associated to vacua of the 3d theory in flat space. The poles in \eqref{polepositions} that contribute to the contours correspond to supersymmetric vacua of the 3d gauge theory in $\Omega$-background along the D3 brane, depending on $q$. 

Rather than giving detailed examples of contours (which one can find in \cite{ACHS, AHS}), let us describe the geometry behind them. 
In a vacuum of the 3d gauge theory, bifundamentals and chiral matter fields get expectation values, due to FI terms which are turned on. This has a geometric interpretation which is very useful \cite{CKV}.  Giving expectation values to bifundamentals corresponds to binding the D3 branes into 2-cycles whose homology classes in $H_2(X, {\mathbb Z}) =\Lambda$ are positive roots $e_{\gamma}$. The restriction to positive roots comes from supersymmetry, which requires us to take positive integer combinations of D3 branes $e_{\gamma} = \sum_a \gamma_a\, e_a $, with $\gamma_a\geq 0$.  In the vacuum with FI terms, chiral multiplets from D3-D5 strings must also have expectation values; this corresponds to ending the D3 branes on a D5 brane which is wrapping a cycle $S_i^*$ in ${\cal W}_{\cal S}$. The 2-cycles  $D_{\gamma, i, \alpha}$ one gets in this way have homology classes corresponding to positive root vectors $e_\gamma$ and have a boundary on a D5 brane wrapping $\omega_i =[S^*_i]$. Depending on a sign of the Fayet Iliopolous term, only the chiral multiplets in fundamental representation can get expectation values. The supersymmetric vacuum describes distributing the D3 branes between the cycles $D_{\gamma, i, \alpha}$, satisfying the conditions that all the wrapping numbers are positive, and that D3 brane charge is conserved. (The index $\alpha$ allows for the possibility that there can be more than one such cycle. We will see such examples in section \ref{sec:Examples}.)
A chiral multiplet that gets expectation value is reflected in poles at points on the Coulomb branch where it becomes massless. In the $\Omega$-background along the D3 branes, one gets families of such vacua shifted by the values of the 3d vortex fluxes along the D3 branes, so the vacua come in families, parametrized by the lengths of the rows of 2d partitions, one for each 3d Coulomb modulus $x$. For the $A_n$ and $D_n$ examples, we have checked explicitly that there is indeed a choice of contours reflecting this structure, and such that evaluating the integral by residues leads to \eqref{3dparta}, with poles at \eqref{polepositions},\eqref{Coulomb}.

\section{Examples}
\label{sec:Examples}

We will now illustrate the general results of previous sections with the basic case of the little string theory on a sphere with three full punctures, in the terminology of \cite{G2}. This corresponds to a ($q$-deformed) 3-point function for Toda CFT,  with 3 primary operator insertions of generic momenta.  In little string theory on a Riemann surface 
${\cal C}$  which is a cylinder, there are two punctures to begin with. The third puncture is introduced by D5 brane defects. 
 
To construct the full puncture defect, as explained in section \ref{subsubsec:WS}, one picks a collection of $n+1$ weight vectors ${\cal W}_{\cal S} = \{ \omega_i\}_{i=1}^{n+1}$ satisfying the following properties: $\omega_i$'s are chosen from the Weyl orbits of the $n$ fundamental weights of ${\bf g}$ (or more precisely, of minus the fundamental weights), they provide a basis of the weight lattice $\Lambda_*$, 
and sum up to zero
\beq\label{fulldefect}
\sum_{i=1}^{n+1} \omega_i =0.
\eeq
From the data given we can find the three theories related by triality, as we explained in section \ref{sec:Triality}:
the 5d gauge theory, the D3 brane gauge theory and the collection of vertex operators in the $q$-deformed Toda CFT they correspond to.

The parameters of the theory are as follows: for each $\omega_i$, we pick a point on ${\cal C}$ with coordinate $x_i=Rm_i.$  $x_i$ encodes the position of the D5 brane wrapping $\omega_i = [S_i^*]$ on ${\cal C}$, and the masses $m_i$'s of matter fields in the 5d and the 3d gauge theory. From Toda perspective, they correspond to $n$ momenta and the the position of the puncture. With only 3 punctures, the latter  can be set to $z=1$. In addition we need to specify $n$ moduli of $(2,0)$ theory we called $\tau_a$'s in \eqref{taua}. They determine the 5d gauge couplings, the 3d FI parameters and correspond to the momentum  of the puncture at $z=0$ from the Toda perspective. There are $n$ more parameters to specify: the $n$ non-normalizable Coulomb moduli associated with the $U(1)$ centers of the 5d gauge groups in \eqref{5dgaugegroup};  the net ranks $N_a$ of the 3d gauge groups in \eqref{3dgauge}, and the numbers of screening charges in \eqref{expect} in Toda theory, or equivalently, the momentum of the puncture at $z=\infty$.

\vskip 0.5cm
{\it 5d Gauge Theory}
\vskip 0.5cm
We split each of the weights  $\omega_i $ into $-w_a$ plus a sum of simple roots with non-negative coefficients. Here, $w_a$ is the fundamental weight in whose orbit $-\omega_i$ lies. This splitting is unique, due to a well known theorem in the theory of Lie algebras and their representations.\footnote{We thank Chrisitan Schmidt for collaboration relating to this point.} This corresponds to splitting of the cycles  wrapped by non-compact D5 branes on the Higgs branch, into a canonical non-compact part and the remaining collection of compact cycles. We collect all the non-compact cycles into  $[S^*] = -\sum_{a=1}^n m_a w_a$, where $m_a$ counts how many $\omega$'s came from the orbit of $-w_a$, and the compact cycles into $[S] = \sum_{a=1}^n d_a e_a$.  As a consequence of  \eqref{fulldefect}, these satisfy the constraint $\sum_{b} C_{ab} d_b = m_a$. This leads to a 5d ${\cal N}=1$ ADE quiver gauge theory with $U(d_a)$ gauge group and $m_a$ fundamental hypermultiplets on the $a$'th node, together with the bifundamental matter fields making up the Dynkin diagram of the Lie algebra. From the quiver, we can compute the partition function bulk theory as the Nekrasov partition function of the 5d gauge theory, following the prescription in section \ref{subsec:5dpart}.

\vskip 0.5cm
{\it 3d Gauge Theory}
\vskip 0.5cm
We take D3 branes in class $[D] = \sum_{a=1}^n N_a e_a$. The 3-3 strings lead to the ${\cal N}=4$  quiver theory with gauge group $\prod_{a=1}^n U(N_a)$ and bifundamental hypemultiplets according to the Dynkin diagram of ${\bf g}$. The ${\cal N}=2$ matter comes from 3-5 strings: strings stretching between the D5 brane on $S^*_i$ and the D3 branes on $S_a$ give $(\omega_i,e_a)$ anti-chiral minus the chiral multiplets in fundamental representation of the  $U(N_a)$ gauge group.  From the quiver, we can compute the partition function of the 3d gauge theory, following the prescription in section \ref{subsec:3dpart}. The partition function depends also on the relative splitting of the $N_a$'s between the vacua. For each positive root $e_{\gamma}$, we get as many vacua as $\omega_i$'s  with $(e_{\gamma}, \omega_i)<0$, counted with multiplicity $|(e_{\gamma}, \omega_i)|$.  This number, as we'll see, always equals the number of Coulomb moduli of the 5d gauge theory. 
\vskip 0.5cm
{\it Toda CFT}
\vskip 0.5cm
The collection of weights $\omega_i$ satisfying \eqref{fulldefect}
leads to the $q$-deformed primary vertex operator $V_{\alpha_1}(z) $
\beq\label{q-deformedV}
 :\prod_{i=1}^{n+1}V_{\omega_i}(x_i):  \qquad \rightarrow \qquad V_{\alpha_1}(z) 
\eeq
where $x_i$, $z$, and $\alpha$, are related as in \eqref{cftposm}, $e^{x_{i}} = z\;q^{ \alpha_{i}}$. With only 3 punctures, no physical quantity will depend on $z$ itself, so we can set it to $1$. The vertex operator $V_{\omega_i}(x)$ corresponding to the weight $\omega_i$ is constructed in appendix $A$. This vertex operator is the $q$-deformation of the primary $V_{\alpha_1}(z)$. Taking the vertex operators to $z=0, \infty$, the details of the $q$-deformation are not important since only the zero modes survive and these are already encoded in $\tau_a$ and $N_a$. The general form of the vertex operator is given in the appendix; specific operators for all the ADE groups will be given below. The vertex operators $V_{\omega_i}(x_i)$ encode the R-charges of the 3d chiral multiplets coming from strings with one end-point on the D5 brane wrapping $S_i^*$, and we can read off from them the subtle $v$ factors in \eqref{Coulomb}.

In Toda CFT, the 3-point function of ${\cal W}$-algebra primaries
\beq\label{3pt}
\langle V_{\alpha_0}(0) V_{\alpha_1}(1) V_{\alpha_{\infty}}(\infty)\rangle
\eeq
is labeled by the three momenta $\alpha_{0}, \alpha_1, \alpha_{\infty}$. If $\alpha_{\infty} = -\alpha_0-\alpha_1 -\sum_{a=1}^n N_a e_a/b$ for positive integers $N_a$ (the ranks of D3 brane gauge theory) we can compute 
the three-point function \eqref{3pt} in free field formalism where we insert $N_a$ screening charge generators
\beq
 \label{3point}
\langle V_{\alpha_0}(0) V_{\alpha_1}(1) V_{\alpha_{\infty}}(\infty) \prod_{a=1}^n Q_a^{N_a}\rangle_{free}.
\eeq
Replacing the vertex operators and the screening charges $Q_a = \int dx \,S_a(x)$ by the $q$-deformed ones, we get the $q$-deformed 3-point conformal block of ${\cal W}_{q,t}(\bf g)$ algebra, as described in section \ref{sec:Toda} and appendix A. The $q$-deformed block is the partition function of the gauge theory on D3 branes given by the corresponding data. In particular, the number of $q$-deformed blocks of the ${\cal W}_{q,t}(\bf g)$ algebra is the number of branches of the 3d gauge theory. Computing \eqref{3point} by residues gives the bulk partition function, as we explained in sec. \ref{sec:Triality}. 

Below, we will make this explicit for all the ADE groups, beginning with the familiar ${\bf g}=A_n$ case, where we will recover the results obtained in \cite{AHS} using a different, $T$-dual setting. We will move on to ${\bf g}=D_n, E_n$  theories, where the technique of the present paper are indispensable. Before we do that, let us digress and discuss the CFT limit of the theory.

\vskip 0.5cm
{\it Coulomb Moduli, Contours and Fusion Multiplicities}

\vskip 0.5cm

Since the description of defects in little string is new, and the $q$-deformed Toda CFT's are not familiar to many, we will pause to demonstrate that in the $m_s$ to infinity limit the 5d gauge theory describes the $(2,0)$ CFT on the 3-punctured sphere, and show how the correspondence to the Toda CFT 3-point conformal block emerges. The D5 branes source the Higgs field of the $(2,0)$ little string theory which, on the Higgs branch, takes the form \eqref{Higgsa}
\beq\label{Higgsa}
e^{R' \varphi_{\cal S}(x)} = e^{\tau} \prod_{{\omega}^{\vee}_i \in {\cal W}_{{\cal S}} }V_i(x)^{{\omega}^{\vee}_{i}}.
\eeq
We defined\footnote{The exponent in $\omega_i^{\vee}$ is to remind us to interpret them as linear combinations of the Chevalley generators of the Cartan subalgebra, see section 2.}
$$V_i(x) = (1-e^{x- m_i R' })^{-1},$$
see sections 2.2.2-3. In the $m_s$ to infinity limit  we described in \ref{subsubsec:defectlimit}, the above becomes
\beq\label{3pole}
\varphi_{S}(z) = {\alpha_0 \over z} +  { \alpha_1 \over 1-z},
\eeq
where
$$
\alpha_1 = \sum_{i=1}^{n+1} m_i\, \omega_i^{\vee}, \qquad \alpha_0 = \tau/R',
$$
and $z=e^{-x}$. This makes it manifest we have $(2,0)$ theory with three punctures on ${\cal C}$. The punctures at $z=0, \infty$ are there because ${\cal C}$ is a cylinder and $\varphi_{\cal S} = \varphi_{S}(z) dz$ a one form on it, with values in the Lie algebra; the puncture at $z=1$ comes from D5 branes. The residues $\alpha_0$ and $\alpha_1$
are both generic elements of the Cartan subalgebra of ${\bf g}$. For $\alpha_1$, this is the case since the $n+1$ weights in ${\cal W}_{\cal S}$ satisfy a single algebraic relation (they sum up to zero) and the D5 brane positions are taken to be generic as well. For $\alpha_0$, this is true because we needed all $n$ gauge couplings $\tau_a$ to be arbitrary complex numbers with ${\rm Re}(\tau_a)>0$.  
From the bulk perspective, the momentum at infinity $\alpha_{\infty}$ is determined by the $n$ non-normalizable Coulomb moduli associated with the $n$ $U(1)$ centers of the $U(d_a)$ gauge groups: the $U(1)$ gauge fields are frozen by the Green-Schwarz mechanism, and the Coulomb moduli that come with them are parameters, not moduli. 

As we explained in section \ref{sec:Toda}, to fully specify the conformal block, we need to specify $h_+({\bf g})-n$ more parameters coming from fusion multiplicities. This matches the number of Coulomb moduli of the bulk theory: we explained in section \ref{subsubsec:defectlimit} that one finds $h_+({\bf g})-n$ normalizable Coulomb. This is also the same as the number of types contours of integration of the D3 brane theory in the limit, or equivalently, the number of choices of vacua of the theory. In \cite{CKV,  Cachazo:2001sg}, the number of vacua of an ADE quiver gauge theory with superpotential  ${\rm Tr} \,{W}_a(X_a)$ for the adjoint chiral multiplet, was studied. It was shown that, if $W_a(x)$ has an isolated critical point for each $a$, there are $h_+({\bf g})-n$ of vacua corresponding to different breaking patterns of the $\prod_{a=1}^n U(N_a)$ gauge group. In a vacuum, the gauge group is broken to $\prod_{\gamma>0} U(N_{\gamma})$ where $e_{\gamma}$ are positive roots. One has $h_{+}({\bf g})$ integers to specify, the number of positive roots, but there are $n$ constraints since this has to come from the original gauge group:  $\sum_{\gamma>0} N_{\gamma}\, e_{\gamma} = \sum_{a=1}^n N_a \, e_a.$
This problem relates to ours by thinking of eigenvalues of $X_a$ as parameterizing the positions of the D3 branes on ${\cal C}$, and integrating out the chiral ${\cal N}=2$ matter to get ${\rm Tr}\, {W}_a(X_a)$ as the effective superpotential. In our context, is easy to show that the effective superpotential satisfies $\partial_x W_a(x) = (e_a, \varphi_S(x))$ \cite{NS, NS1, ASv}, and has one critical point for each $a$. 
To establish more detailed correspondence with the D3 brane theory or to Toda CFT, one has to go back to the little string, since no other way is known to compute the partition function of the $(2,0)$ CFT on ${\cal C}$.


After the $q$-deformation, the number of Coulomb branch moduli of the bulk theory, the vacua of the 3d gauge theory on D3 branes, and the contours of integration of $q$-deformed conformal blocks are equal to each other in all ADE theories, but they are larger than $h_+({\bf g})-n$, except for ${\bf g}=A_n$. As we explained earlier, this is not surprising from neither the bulk nor the Toda perspective, as the massive theory contains more data than its CFT limit. In 5d gauge theory, the number of normalizable Coulomb moduli is simply $\sum_{a=1}^n (d_a-1)$. From the 3d gauge theory perspective, counting vacua is more complicated, and it is remarkable that one always finds agreement. The vacua are still related to counting positive roots, but now with multiplicity, as we explained in section \ref{subsec:Equality}. In the vacuum, the D3 brane gauge group is broken to 
$\prod_{D_{\gamma, i, \alpha}} U(N_{\gamma, i, \alpha})$ where we get a gauge group factor for each cycle $[D_{\gamma, i, \alpha}]=e_{\gamma}$ with boundary on $[S_i^*] =\omega_{i}$. We get $|(e_\gamma, \omega_i)| = I_{\gamma,i}$ of such cycles if $(e_a, \omega_i)$ is negative -- we used the index $\alpha$ to label them. If $(e_a, \omega_i)$ is not negative, we get none. Thus, to specify the breaking pattern, we get to pick an integer for each of the positive roots $e_{\gamma}$, counted with multiplicity $\sum_{\omega_i \in {\cal W}_{\cal S}} I_{\gamma,i}$. The integers have to satisfy $n$ constraints for the total ranks to add up to $N_a$, for each $a$. This count also equals the number of choices in specifying the contour of integration in \eqref{3dpart}, and the number of fusion multiplicities of ${\cal W}_{q,t}({\bf g})$ algebra, since the same integral computes both \eqref{3dpart} and \eqref{3point}, after $q$-deformation. 

\subsection{$A_n$ Little String Theory}
For the $n+1$ weight vectors $\omega_i = [S_i^*]$ we choose:
\beq
 \begin{aligned}\label{fund}
&\omega_1=- {w}_1, \\
&\omega_2 = -w_1+e_1,\\
&  \vdots\\ 
&\omega_{n} = -w_1+e_1+\ldots + e_{n-1},\\
&\omega_{n+1} = -w_1+e_1+\ldots+ e_{n-1}+e_n.
\end{aligned}
\eeq
Here,
$w_1$ is the highest weight of the $1$st-fundamental representation; the rest of the weights on the right hand side of \eqref{fund} are obtained from it by acting with the Weyl group of $A_n$; the weights $-\omega_i$ are the $n+1$ weights of the defining, $n+1$ dimensional representation of $A_n$ ("the fundamental representation").  An equivalent way to write \eqref{fund} is in terms of fundamental weights, using $e_a = \sum_{b=1}^n C_{ab} \, w_b$:
\beq
 \begin{aligned}\label{funda}
&\omega_1=- {w}_1, \\
&\omega_i = -{w}_i + w_{i-1},  \qquad i=2, \ldots n,\\
&\omega_{n+1} = w_n
\end{aligned} 
\eeq
The set of weights ${\cal W}_{\cal S} = \{\omega_i\}_{i=1}^{n+1}$ spans the weight lattice and satisfies \eqref{fulldefect}. \subsubsection{The 5d $A_n$ Gauge Theory}
To find 5d gauge theory description of the $(2,0)$ little string, we sum up the simple roots on the right and side of \eqref{fund} to get $[S]$, the homology class of D5 branes wrapping compact 2-cycles:
$$
[S] = n e_1+ (n-1) e_2+\ldots + e_{n}
$$
This leads to an $A_n$ quiver theory where the coefficient of $e_a$ in $[S]$ is the rank of the gauge group on the $a$-th node, $d_a =n-a+1$. The fundamental matter
comes from non-compact D5 branes whose homology class $[S_*]$ is the sum of (minus) the fundamental weights in \eqref{fund}
$$[S_*] = -(n+1) w_1.
$$
To number of hypermultiplets transforming in fundamental representation of the gauge group on the $a$-th node is the coefficient of $-w_a$ in the expansion of $[S_*]$ in terms of fundamental weights. In this case,  we have $n+1$ hypermultiplets in fundamental representation of the gauge group on the first node. Altogether, we get the 5d ${\cal N}=1$ quiver gauge theory in Fig. \ref{fig:Anfigure}. This agrees with the $T_N$ theory shown in \cite{AHS} to describe the 3-point function of $A_n$ with $N=n+1$, using a dual string realization. Given the lagrangian description of the low energy theory, the partition function can be obtained using the prescription in section \ref{subsec:5dpart}.
The 5d $T_N$ theory was also obtained in \cite{Bergman:2014kza, Hayashi:2014hfa}.
\subsubsection{The 3d $A_n$ Gauge Theory}
\begin{figure}
\emph{}
\hspace{-7ex}
\centering
 \vspace{-30pt}
\includegraphics[width=0.50\textwidth]{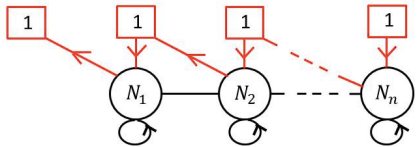}
 \vspace{-30pt}
\caption{D3 quiver from $A_n$ little string.} 
\label{fig:AnQuiver}
\end{figure}

We start with the collection of D5 branes wrapping $n+1$ non-compact cycles $S_i^*$ with $[S_i^*]=\omega_i$, arising in the Higgs phase of the 5d gauge theory. We add the D3 branes wrapping the  compact two-cycles in the homology class of $D$:
$$
[D] = \sum_{a=1}^n N_a \, e_a. 
$$
The strings beginning and ending on D3 branes lead to the ${\cal N}=4$ $A_n$ quiver theory with gauge group $\prod_{a=1}^n U(N_a)$ and bifundamental hypermultiplets according to the $A_n$ Dynkin diagram. The ${\cal N}=2$ matter comes from intersections of D3 branes with D5 branes in ${\cal W}_{\cal S}$.  The number of chiral minus the anti-chiral multiplets in fundamental representation of the gauge group on the $a$-th node, coming from a D5 brane on $\omega_i$, is
$$
\#(S_a, S_i^*) = (e_a, \omega_i) = -\delta_{a,i}+\delta_{a, i-1},
$$
as we explained in section \ref{subsec:3dgauge}. This is the coefficient of $w_a$ in the expansion of $\omega_i$ in terms of the fundamental weights in \eqref{funda}. The theory on the D3 branes is an $A_n$ ${\cal N}=2$ quiver theory in Fig. \ref{fig:AnQuiver}.
This agrees with the quiver of vortex theory derived in \cite{AHS} using different means. From the quiver, we can compute the partition function of the 3d gauge theory, following the prescription in section 3.2.

%

\subsubsection{$A_n$ Toda Conformal Block}
The $q$-deformed vertex operator is $:\prod_{i=1}^{n+1} V_{\omega_i}(x_i):$ where
\beq
 \begin{aligned}\label{fundv}
&V_{\omega_1} (x)= W_1^{-1}(x), \\
&V_{\omega_2} (x)= : W_1^{-1}(x) E_1(xv^{-1}):\\
&  \vdots\\ 
&V_{\omega_n} (x)= : W_1^{-1}(x) E_1(xv^{-1}) E_2(x v^{-2}) \ldots  E_{n-1}(xv^{1-n}):\\
&V_{\omega_{n+1}} (x)= : W_1^{-1}(x) E_1(xv^{-1}) E_2(x v^{-2}) \ldots  E_{n-1}(xv^{1-n}) E_n(x v^{-n}):
\end{aligned}
\eeq
in terms fundamental weight and simple root vertex operators in the appendix A. This way of writing the vertex operators encodes the subtle $v$ dependence of the Coulomb moduli at the triality point in \eqref{Coulomb}. The vertex operators can be thought of as quantizing the classical formula \eqref{fund}. We have 
$e_{(a), I} = f_i \,t^{N_{(a), I}} \, v^{-2a+1}$, where $I$ runs from $1$ to $n-a+1$. This comes with a factor $v^{-a}\times v^{-a+1}=v^{-2a+1}$, independently of which $\omega_i$  the simple root $e_a$ gets assigned to. The factor $v^{-a}$ comes from \eqref{fundv} and is the subtle one; the second factor $v^{\#_a} =v^{-a+1}$ can be read off from the Dynkin diagram as defined in section \ref{subsec:Equality}. This reflects the ${\cal W}_{q,t}(A_n)$ algebra symmetry of the theory in $\Omega$ background.

There is an equivalent way of writing the vertex operators \eqref{fundv}:
\beq
 \begin{aligned}\label{funda}
&V_{\omega_1} (x)=W_1^{-1}(x), \\
&V_{\omega_i} (x) =W_i^{-1}(x v^{-i+1}) W_{i-1}(xv^{-i}), \qquad i=2, \ldots n,\\
&V_{\omega_{n+1}} (x) =  W_{n}(xv^{-n-1}).
\end{aligned} 
\eeq
Written this way, the vertex operators encodes the $R$-charges of chiral multiplets. The $U(1)_R$-charge $S_R =-r/2$ of an (anti)chiral multiplet from 5-3 strings ending on D5 brane in class $\omega_i$ and the D3 brane on the $a$'th node is encoded in the $v$-shift of the argument of $W_{a}(xv^{r})$.
The vertex operators, essentially in this form, appeared in \cite{FR1} in section 5.1.1, with obvious translations.

The number of normalizable Coulomb moduli of the 5d theory is $\sum_{a=1}^n(d_a-1) = h_+(A_n)-n$, so the dimension of the Coulomb branch is the same in the little string and in the $(2,0)$ CFT.  This also matches the number of vacua/contours of integration in \eqref{3dpart}, which can be counted using the prescription in section \ref{subsec:Equality}: The positive roots $e_{\gamma}$ of the $A_n$ Lie algebra are of the form $e_{i}+e_{i+1}+\ldots + e_j$, for $i<j$. Each positive root contributes with multiplicity $1$, since each root has negative multiplicity with exactly of the weights in ${\cal W}_{\cal S}$, and the intersection equals to $-1$. For the root we wrote, the corresponding weight is ${\cal \omega}_i$. Thus the breaking pattern of $\prod_{a=1}^n U(N_a)$ allows for $h_+(A_n)-n$ choices. This is the same as the data needed to specify the $q$-deformed ${\cal W}_{q,t}(A_n)$ algebra 3-point block. As we explained in \ref{sec:Toda} this number could have been greater, in principle, than the number of 3-point functions of the $A_n$ Toda CFT.  In this case, it is not: The ${\cal W}(A_n)$ algebra symmetry is generated by currents of spins $s_a=2,3,\ldots, n+1$ leading to $\sum_{a=1}^n (s_a-2) ={1\over 2} {n(n+1)}-n=h_+(A_n)-n$ additional parameters needed to specify the conformal block.
\subsection{$D_n$ with $3$ Full Punctures}
For ${\cal W}_{\cal S}$, we take the following collection of $n+1$ weights of $D_n$
\beq
 \begin{aligned}\label{fundwd}
&\omega_1=- {w}_1 + e_1+e_{2} + \ldots +e_{n-2} + e_{n-1} + e_n,\\
& \omega_i = \omega_{i-1} + e_{n-i}, \qquad i=2, \ldots n-1\\
&  \omega_{n}= -{w}_{n-1} \\
 & \omega_{n+1} = -{w}_n 
\end{aligned}
\eeq
where $w_a$ is the $a$'th fundamental weight. The weights $w_1$, $w_{n-1}$, and $w_{n}$, are respectively the highest weights of the $2n$ dimensional defining representation of $D_n$ and the two $2^{n-1}$ dimensional spinor representations. 
The first set of $n-1$ weights in \eqref{fundwd} are all in the orbit of $w_1$. Another way to write \eqref{fundwd} is as follows:
\beq
 \begin{aligned}\label{fundwda}
&\omega_1=- {w}_{n-2}+w_{n-1} + w_n, \\
&\omega_i = -w_{n-i-1} + w_{n-i}, \qquad i=2, \ldots n-1,\\
&\omega_{n} = -w_{n-1},\\
&\omega_{n+1} = -w_n
\end{aligned} 
\eeq
This way of writing $\omega$'s makes it easy to check that the set ${\cal W}_{\cal S}$ spans the weight lattice, and that $\omega_i$'s sum up to zero.

\subsubsection{The 5d $D_n$ Gauge Theory}

To find the 5d gauge theory description of the $(2,0)$ little string, we sum up the simple roots on the right and side of \eqref{fundwd}:
$$
[S] = n e_1+ (n+1) e_2+\ldots + (2n-3) e_{n-2} +(n-1) e_{n-1} +(n-1) e_n.
$$
The rank of the $a$'th gauge group in the $D_n$ quiver diagram is the coefficient of $e_a$. The fundamental matter
comes from non-compact D5 branes whose homology class $[S_*]$ is the sum of (minus) the fundamental weights in \eqref{fundwd}
$$[S_*] = -(n-1) w_1 -w_{n-1}-w_n.
$$
The number of fundamental hypermultiplets on the $a$-th node is computed by the coefficient of $-w_a$ in $[S_*]$.
We get $n-1$ hypermultiplets in fundamental representation of the gauge group on the first node, and one each for the $n-1$'st and the $n$'th node. The resulting 5d gauge theory has a quiver diagram given in figure Fig.\ref{fig:Dnfigure}. It satisfies the condition $\sum_{b} C_{ab}\, d_b = m_a$.
From the quiver, the partition function of the $(2,0)$ theory follows immediately, applying the formalism of section \ref{subsec:5dpart}.



\subsubsection{The 3d $D_n$ Gauge Theory}

We take the D3 branes to wrap the collection of compact two-cycles in the homology class of $[D] = \sum_{a=1}^n N_a \, e_a.$
The ${\cal N}=2$ matter comes from intersections of D3 branes in class $e_a$ with D5 branes in class $\omega_i$; the number of anti-chiral multiplets minus the chiral ones is the coefficient of $w_a$ in the expansion of $\omega_i$ in \eqref{fundwda}. 
\begin{figure}
\hspace{-7ex}
\centering
 \vspace{-20pt}
\includegraphics[width=0.55\textwidth]{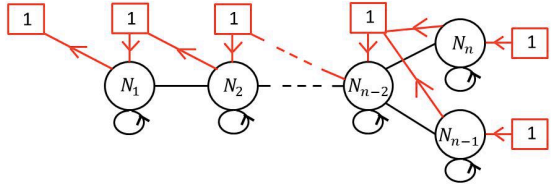}
 \vspace{-20pt}
\caption{D3 quiver from $D_n$ little string.} 
\label{fig:DnQuiver}
\end{figure}
This leads to the 3d quiver in figure
\ref{fig:DnQuiver}.
From the quiver, we can compute the partition function of the 3d gauge theory, following the prescription in section 3.2.


\subsubsection{$D_n$ Toda conformal block}

Corresponding to weights in \eqref{fundwd} we get the vertex operator $:\prod_{i=1}^{n+1} V_{\omega_i}(x_i):$, with
\beq
 \begin{aligned}\label{fundedv}
&V_{\omega_1} (x)= :W_1^{-1}(x)E_1(xv^{-1})E_2(x v^{-2}) \ldots  E_{n-2}(x v^{-n+2})E_{n-1}(x v^{-n+1})E_{n}(x v^{-n+1}):, \\
&V_{\omega_i} (x)= : V_{\omega_{i-1}}(x)E_{n-i}(x v^{-n-i+2}):, \qquad i=2, \ldots n-1 \\
&V_{\omega_n} (x)= : W_{n-1}^{-1}(x) :\\
&V_{\omega_{n+1}} (x)= : W_{n}^{-1}(x), :
\end{aligned}
\eeq
in terms fundamental weight and simple root vertex operators in the appendix A. This way of writing the vertex operators encodes the subtle $v$ dependence of the Coulomb moduli at the triality point in \eqref{Coulomb}. The vertex operators can be thought of as quantizing the classical formula \eqref{fund}. Each Coulomb modulus $e_{(a),I}$ gets assigned to one and only one simple root vertex operator $E_a$ in \eqref{fundedv}. This map also encodes the $\omega_i$ to which the simple root $e_a$ gets assigned to. We can read off that $e_{(a),I} = f_i \,t^{N_{a,I}} v^{\#_{a,i,I}} $, where  $v^{\#_{a,i,I}}$ is equal to $v^{\#'_{a,i,I}}$, which enters the argument of the vertex operator $E_a$, times $v^{\#a}$ defined in section \ref{subsec:Equality}. The power of $v$ in $v^{\#_a}$ counts the position of the $a$-th node in the Dynkin diagram, in terms of number of links that separate it from the first node, so that $v^{\#_a} = v^{1-a}$, for $a$ between $1$ and $n-2$, and $v^{\#_{n-1}}= v^{\#_{n}} = v^{2-n}$.

The vertex operators can also be written in terms of fundamental weight vertex operators alone, using results from the appendix A. This gives
\beq
 \begin{aligned}\label{fundedv}
&V_{\omega_1} (x)= :W_{n-2}(x v^{-n+1})^{-1} W_{n-1}(x v^{-n}) W_{n}(x v^{-n}): \\
&V_{\omega_i} (x)= : W_{{n-i-1}}^{-1}(xv^{-n-i+2})W_{n-i}(x v^{-n-i+1}):, \qquad i=2, \ldots n-1 \\
&V_{\omega_n} (x)= : W_{n-1}^{-1}(x) :\\
&V_{\omega_{n+1}} (x)= : W_{n}^{-1}(x) :
\end{aligned}
\eeq
This way of writing the vertex operators encodes the $U(1)_R$ symmetry charges of chiral multiplets, in the same manner as in the $A_n$ case. This should correspond to the 3d theory being conformal in the IR, although we have not attempted to check that. The vertex operators $V_{\omega_i}$, coming from the fundamental representation of $D_n$, with highest weight $w_1$, are written, essentially in this form, in \cite{FR1} in section 5.1.4 of the paper.

The $D_n$ Toda CFT has ${\cal W}(D_n)$ algebra symmetry generated by currents of spins $s_a=2,4,\ldots, 2(n-1), n$. To fully specify the 3-point conformal block of the Toda CFT in \eqref{3point}, one has to specify the additional $\sum_{a=1}^n (s_a-1) =n^2 - n= h_+(D_n) $ parameters, besides $\alpha_0$ and $\alpha_1$. This is also the number of Coulomb moduli of the bulk theory and the D3 brane vacua in the limit where $m_s$ goes to infinity. In the little string case, all these numbers are larger:
The number of Coulomb moduli of the bulk theory is $\sum_{a=1}^n (d_a-1) =\dfrac{1}{2}(n-1)(3n-2) -n $. This is also the number of integers we get to pick to specify the vacuum of the 3d theory, obtained by counting positive roots, with multiplicity:
$\dfrac{(n-1)(n+2)}{2}$ positive roots have come with multiplicity $1$ as they have intersection number $-1$ with one of the weights, 
and $\dfrac{(n-1)(n-2)}{2}$ positive roots count with multiplicity $2$, as they have intersection number $-1$ with two of the weights. Fortunately,
$\dfrac{(n-1)(n+2)}{2}+ 2\cdot \dfrac{(n-1)(n-2)}{2}=\dfrac{1}{2}(n-1)(3n-2)$, leading to the same count as the Coulomb moduli. 

\subsection{$E_6$ with $3$ Full Punctures}

For the weight system ${\cal W}_{\cal S}$, we take:
\beq\label{funde6}
\begin{aligned}
 & \omega_1 = -{w}_5 \\
 & \omega_2= -{w}_5 + e_5\\
& \omega_3= -{w}_5 + e_1 + 2e_2 + 3e_3 + 3e_4  + 2e_5 + 2e_6\\
& \omega_4= -{w}_5 + e_1 + 2e_2 + 4e_3 + 3e_4  + 2e_5 + 2e_6 \\
& \omega_5= -{w}_5 + e_1 + 3e_2 + 4e_3 + 3e_4  + 2e_5 + 2e_6 \\
& \omega_6= -{w}_5 + 2e_1 + 3e_2 + 4e_3 + 3e_4  + 2e_5 + 2e_6 \\
& \omega_7= -{w}_6 
\end{aligned}
\eeq 
The first 6 weight vectors in \eqref{funde6} can be seen to lie in the Weyl orbit of $-w_5$. This can be rewritten as
\beq\label{funde6a}
\begin{aligned}
 & \omega_1 = -{w}_5 \\
 & \omega_2= -{w}_4 + w_5\\
& \omega_3= -{w}_3+w_4 + w_6\\
& \omega_4= -{w}_2+w_3\\
& \omega_5= -{w}_1+w_2  \\
& \omega_6= {w}_1 \\
& \omega_7= -{w}_6 
\end{aligned}
\eeq 
which makes it easy to check that ${\cal W}_{\cal S}$ provides a basis of the weight lattice of $E_6$ and the weights in ${\cal W}_{\cal S}$ sum up to zero.

\subsubsection{The Bulk $E_6$ Gauge theory}
Summing up the fundamental weight and the simple root part of the right hand side in \eqref{funde6} separately, we find
\beq\label{e6g}
[S] = 5 e_1+10 e_2 + 15 e_3+ 12 e_4+9 e_5 + 8 e_6.
\eeq
and
\beq\label{e6f}
[S_*] = -6 w_5 - w_6
\eeq
This leads to a 5d quiver gauge theory description in figure \ref{fig:E6figure}. The partition function of the theory can be computed from the quiver, as in section \ref{subsec:5dpart}. 

\subsubsection{The D3 Brane $E_6$ Gauge theory}

\begin{figure}
\emph{}
\hspace{-7ex}
\centering
 \vspace{-20pt}
\includegraphics[width=0.55\textwidth]{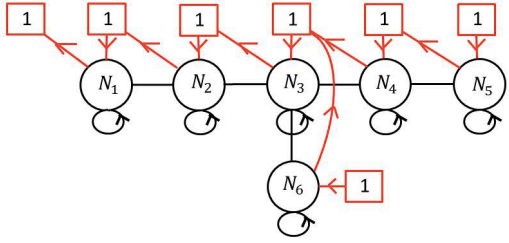}
 \vspace{-20pt}
\caption{D3 quiver from $E_6$ little string.} 
\label{fig:E6quiver}
\end{figure}

The ${\cal N}=2$ matter can be read off from \eqref{funde6a}. This leads to the 3d quiver in figure \ref{fig:E6quiver}.

\subsubsection{$E_6$ Toda CFT}
The E6 vertex operators corresponding to \eqref{funde6} are the following:
\beq
 \begin{aligned}\label{funde6v}
V_{\omega_{1}} (x)=& : W_{5}^{-1}(x) :, \\
V_{\omega_{2}} (x)=& :W_5^{-1}(x)E_5(xv^{-1}):, \\
V_{\omega_{3}} (x)=& :W_5^{-1}(x)E_5(xv^{-1})E_4(xv^{-2})E_3(xv^{-3})E_2(xv^{-4})E_6(xv^{-4})E_1(xv^{-5})\\
&\;\;\;E_3(xv^{-5})E_2(xv^{-6})E_4(xv^{-6})E_3(xv^{-7})E_5(xv^{-7})E_4(xv^{-8})E_6(xv^{-8}):, \\
V_{\omega_{4}} (x)=& :V_{\omega_{3}} (x)E_3(xv^{-9}):, \\
V_{\omega_{5}} (x)=& :V_{\omega_{4}} (x)E_2(xv^{-10}):, \\
V_{\omega_{6}} (x)=& :V_{\omega_{5}} (x)E_1(xv^{-11}):, \\
V_{\omega_{7}} (x)=& : W_{6}^{-1}(x) :
\end{aligned}
\eeq
From this, we can read off the Coulomb moduli in \eqref{Coulomb}: we map $e_{(a), I}$ in one to one way to the $E_{a}$ vertex operators in \eqref{funde6v}, and 
 set $e_{(a), I} = f_i t^{N_{(a), I}}v^{\#_{a, i, I}}$, where $f_i$ corresponds to the $\omega_i$ to which we assign this simple root. The $v$-factor is encoded in the vertex operator \eqref{funde6v}, we just need to shift the power of $v$ in the argument of the vertex operator by $v^{\#_a}$.
 
Rewriting \eqref{funde6v} in terms of the fundamental vertex operators, using the relations \eqref{etow},
\beq
 \begin{aligned}\label{fundeE6vertex}
& V_{\omega_{1}} (x)= :W_{5}^{-1}(x) :, \\
& V_{\omega_{2}} (x)= :W_5(xv^{-2})W_4^{-1}(xv^{-1}):, \\
& V_{\omega_{3}} (x)= :W_4(xv^{-9})W_6(xv^{-9})W_3^{-1}(xv^{-8}):\\
& V_{\omega_{4}} (x)= :W_3(xv^{-10})W_2^{-1}(xv^{-9}):, \\
& V_{\omega_{5}} (x)= :W_2(xv^{-11})W_1^{-1}(xv^{-10}):, \\
& V_{\omega_{6}} (x)= :W_1(xv^{-12}):, \\
& V_{\omega_{7}} (x)= :W_{6}^{-1}(x) :
\end{aligned}
\eeq
These encode the $U(1)_R$-charges of chiral multiplets. The (anti-)chiral multiplet in fundamental representation of the $a$'th gauge group and coming from the D5 brane on $\omega_i$ has  R-charge $S_R = -r/2$, where $v^r$ is the $v$-dependence in the argument of the vertex operator $W^{\pm1}_a(xv^{r})$ on the right hand side of $V_{\omega_i}(x)$ in \eqref{fundeE6vertex}.

The number of positive roots of $E_6$ is $h_+(E_6) = 36$. The spins of the generators of ${\cal W}(E_6)$ algebra are $s_a = 2,5,6,8,9,12.$  In the ${\cal W}(E_6)$ algebra, to specify the 3-point block, we need $\sum_{a=1}^6 (s_a-2) = 36-6$ parameters besides $\alpha_{0,1,\infty}$.
The number of Coulomb moduli of the 5d theory, including the non-dynamical $U(1)$ factors, is $59-6$. This is also the number of integers we need to specify  the vacuum of the 3d theory: the latter is obtained by counting positive roots with multiplicity. The 36 positive roots of $E_6$ split up into  $19$ positive roots with multiplicity $1$; $13$ positive roots with multiplicity two due to having intersection number $-1$ with two of the weights; $1$ positive root with multiplicity $2$ due to  intersection number $-2$ with one of the weights, $4$ positive roots have intersection number $-1$ with three of the weights. This gives $59$ altogether, and we still have to subtract $6$ for the net rank constraints.

\subsection{$E_7$  with $3$ Full Punctures}
For the weight system ${\cal W}_{\cal S}$, we take:
\beq\label{funde7}
\begin{aligned}
&\omega_1= -{w}_1 + 3e_1 + 5e_2 + 7e_3 + 6e_4  + 4e_5 + 2e_6 + 4e_7\\
&\omega_2 = -{w}_1 + 3e_1 + 5e_2 + 8e_3 + 6e_4  + 4e_5 + 2e_6 + 4e_7 \\
 &\omega_3 = -{w}_1 + 3e_1 + 6e_2 + 8e_3 + 6e_4  + 4e_5 + 2e_6 + 4e_7 \\
& \omega_4 = -{w}_1 + 4e_1 + 6e_2 + 8e_3 + 6e_4  + 4e_5 + 2e_6 + 4e_7\\
&\omega_5 = -{w}_6 \\
&\omega_6 = -{w}_6 + e_6 \\
& \omega_7 = -{w}_6 + e_5 + e_6\\
 &\omega_8 = -{w}_7
\end{aligned}
\end{equation} 
The first four weights are in the Weyl orbit of $-w_1$, the next three in the orbit of $-w_6$. We can rewrite this as
\beq\label{funde7a}
\begin{aligned}
&\omega_1=-w_3+w_4+w_7 \\
&\omega_2 =  -w_2+w_3\\
 &\omega_3 = -w_1+w_2 \\
& \omega_4 =w_1 \\
&\omega_5 =-w_6 \\
&\omega_6 = -w_5+w_6 \\
& \omega_7 = -w_4+w_5\\
 &\omega_8 = -w_7
\end{aligned}
\end{equation} 
from which it is easy to check that ${\cal W}_{\cal S}$ spans the weight lattice and that $\omega_i$'s sum up to zero.

\subsubsection{The Bulk $E_7$ Gauge Theory}
To find the 5d gauge group, we sum up the simple roots on the right hand side of \eqref{funde7} to get: 
\beq\label{e7g}
[S] = 13 e_1 + 22 e_2 + 31 e_3 + 24 e_4 + 17 e_5 + 10 e_6 + 16 e_7,
\eeq
The fundamental weights sum up to
\beq\label{e7f}
[S_*] = -4 w_1 - 3 w_6 - w_7,
\eeq
This leads to the quiver gauge theory in figure \ref{fig:E7figure}, from which we can compute the bulk partition function, using methods of section \ref{subsec:5dpart}.

\subsubsection{The 3d $E_7$ Gauge Theory}
To read off the ${\cal N}=2$ chiral matter content of the 3d quiver gauge theory on the D3 branes, we use the second way of writing the weight system. 
\begin{figure}[h!]
\emph{}
\hspace{-7ex}
\centering
 \vspace{-20pt}
\includegraphics[width=0.55\textwidth]{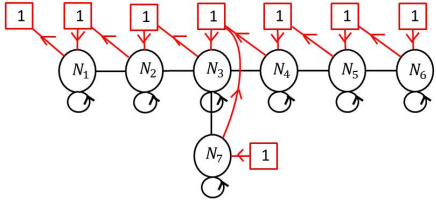}
\vspace{-20pt}
\caption{D3 quiver from $E_7$ little string.} 
\label{fig:E7quiver}
\end{figure}
The number of ${\cal N}=2$ (anti)-chiral multiplets in the fundamental representation of the gauge group on the $a$-th node, coming from D5 branes in class $\omega_i$, is the coefficient of $-w_a$ in the expansion of $\omega_i$ in \eqref{funde7a}. This leads to the quiver in figure \ref{fig:E7quiver}.
\subsubsection{$E_7$ Toda CFT}
We will write the $q$-deformed vertex operators only in terms of the weights this time:
 \begin{align*}
& V_{\omega_{1}} (x)= :W_4(xv^{-15})W_7(xv^{-15})W_3^{-1}(xv^{-14}):, \\
& V_{\omega_{2}} (x)= :W_3(xv^{-16})W_2^{-1}(xv^{-15}):, \\
& V_{\omega_{3}} (x)= :W_2(xv^{-17})W_1^{-1}(xv^{-16}):\\
& V_{\omega_{4}} (x)= :W_1(xv^{-18}):, \\
& V_{\omega_{5}} (x)= :W_6^{-1}(x):, \\
& V_{\omega_{6}} (x)= :W_6(xv^{-2})W_5^{-1}(xv^{-1}):, \\
& V_{\omega_{7}} (x)= :W_5(xv^{-3})W_4^{-1}(xv^{-2}):,\\
& V_{\omega_{8}} (x)= :W_7^{-1}(x) :
\end{align*}
The explicit formulas in terms of simple root vertex operators can easily be obtained from this, but they are too long, because of large numbers of simple roots on the right hand side of \eqref{funde7}. They can be easily obtained from above, using expressions in the appendix.

The number of positive roots of $E_7$ is $h_+(E_7) = 63$. The ${\cal W}(E_7)$ algebra is generated by operators of spins $s_a = 2,6,8,10,12,14,18$. The number of additional parameters, besides the three momenta $\alpha_{0,1, \infty}$ needed to specify the 3-point conformal block $\sum_{a=1}^7 (s_a-2) = 63-7 $. This is the number of normalizable and log-normalizable Coulomb moduli of the bulk theory in the $m_s$ to infinity limit. For finite $m_s$, the number of moduli is larger. The bulk theory has: $133-7$ Coulomb moduli. Tho specify the vacuum of the 3d quiver gauge theory, we need as many parameters as positive roots counted with multiplicities: $25$ roots have multiplicity $1$; $21$ positive roots have multiplicity $2$ as they have intersection number $-1$ with two of the weights; $7$ roots have multiplicity $2$ as they have intersection $-2$ with one of the weights; $16$ roots have multiplicity $3$, as they have intersection number $-1$ with three of the weights; one positive root has intersection number $-1$ with $4$ weights, and hence multiplicity $4$. The count of roots with multiplicities gives $133$, and we have to subtract $7$ constraints on the net ranks. This is the same as the number of Coulomb moduli in the bulk $(2,0)$. It is also the same as the number of choices in assigning contour of integration in the free field formulation of the $q$-deformed 3-point conformal block of the ${\cal W}_{q,t}(E_7)$ algebra.

\subsection{$E_8$ with $3$ Full Punctures}
For the weight system ${\cal W}_{\cal S}$, we take 
\beq\label{funde8}
\begin{aligned}
& \omega_1 = -{w}_1 + 7e_1 + 13e_2 + 19e_3 + 16e_4  + 12e_5 + 8e_6 + 4e_7 + 10e_8\\
 & \omega_2 = -{w}_1 + 7e_1 + 13e_2 + 20e_3 + 16e_4  + 12e_5 + 8e_6 + 4e_7 + 10e_8\\
& \omega_3= -{w}_1 + 7e_1 + 14e_2 + 20e_3 + 16e_4  + 12e_5 + 8e_6 + 4e_7 + 10e_8\\
& \omega_4= -{w}_1 + 8e_1 + 14e_2 + 20e_3 + 16e_4  + 12e_5 + 8e_6 + 4e_7 + 10e_8\\
& \omega_5= -{w}_7 \\
& \omega_6 = -{w}_7 + e_7 \\
 & \omega_7 = -{w}_7 + e_6 + e_7 \\
& \omega_8 = -{w}_7 + e_5 + e_6 + e_7 \\
 & \omega_9= -{w}_8 
 \end{aligned}
\end{equation} 
\noindent
The first $4$ weights $\omega_i$ above are in the Weyl group orbit of $w_1$, the next four in the one of $w_7$. This can be rewritten as:
\begin{align*}
& \omega_1 = -w_3+w_4+w_8 \\
 & \omega_2 =-w_2+w_3 \\
& \omega_3=-w_1+w_2\\
 & \omega_4= w_1\\
& \omega_5= -{w}_7 \\
& \omega_6 = -w_6+w_7 \\
 & \omega_7 = -w_5+w_6\\
& \omega_8 =  -w_4+w_5\\
 & \omega_9= -{w}_8 .
 \end{align*}
\noindent
The weight system provides a basis for the weight lattice of $E_8$, and the $9$ vectors sum up to zero.

\subsubsection{The Bulk $E_8$ Gauge Theory}
The $E_8$ quiver gauge theory describing the low energy dynamics of the bulk $E_8$ little string theory on a sphere with three full punctures is given in figure \ref{fig:E8figure}. 
To derive the quiver, we sum up separately the simple roots and the fundamental weights on the right hand side of \eqref{funde8}. The simple roots sum to
\beq\label{e8g}
[S] = 29 e_1 + 54 e_2 + 79 e_3 + 64 e_4 + 49 e_5 + 34 e_6 + 19 e_7 + 40 e_8,
\eeq
whose coefficients are the ranks of the corresponding gauge groups.
The fundamental weights sum up to
\beq\label{e8f}
[S_*] = -4 w_1 - 4 w_7 - w_8,
\eeq
where the coefficients encode the ranks of the fundamental flavor groups for each node of the quiver.
With the gauge theory description in tow, we can compute the partition function of the bulk little string using results in \ref{subsec:5dpart}.
 \subsubsection{The 3d $E_8$ Gauge Theory}
\begin{figure}[h!]
\emph{}
\hspace{-7ex}
\centering
 \vspace{-25pt}
\includegraphics[width=0.55\textwidth]{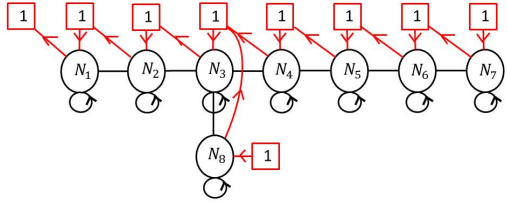}
 \vspace{-20pt}
\caption{D3 quiver from $E_8$ little string.} 
\label{fig:E8quiver}
\end{figure}

The second way of writing the weight system lets us read off the ${\cal N}=2$ chiral matter content of the 3d quiver gauge theory on the D3 branes. The number of ${\cal N}=2$ (anti)-chiral multiplets on the $a$-th node are encoded in the coefficients of $w_a$ in the expansion of $\omega_i$ in terms of fundamental weights. This leads to the quiver in figure \ref{fig:E8quiver}.
From the quiver, we can compute the partition function of the theory as in section \ref{subsec:3dpart}.
\subsubsection{$E_8$ Toda CFT}

The $q$-deformed vertex operator is $\prod_{i=1}^9 V_{\omega_i}(x_i)$ where:

 \begin{align*}
& V_{\omega_{1}} (x)= :W_4(xv^{-27})W_8(xv^{-27})W_3^{-1}(xv^{-26}):, \\
& V_{\omega_{2}} (x)= :W_3(xv^{-28})W_2^{-1}(xv^{-27}):, \\
& V_{\omega_{3}} (x)= :W_2(xv^{-29})W_1^{-1}(xv^{-28}):\\
& V_{\omega_{4}} (x)= :W_1(xv^{-30}):, \\
& V_{\omega_{5}} (x)= :W_7^{6}(xv^{6}):, \\
& V_{\omega_{6}} (x)= :W_7(xv^{4})W_6^{-1}(xv^{5}):, \\
& V_{\omega_{7}} (x)= :W_6(xv^{3})W_5^{-1}(xv^{4}):,\\
& V_{\omega_{8}} (x)= :W_5(xv^{2})W_4^{-1}(xv^{3}):\\
& V_{\omega_{9}} (x)= :W_8^{-1}(xv^{3}) :
\end{align*}

The number of positive roots of $E_8$ is $h_+(E_8)=120$. The ${\cal W}(E_8)$ algebra is generated by operators of spins $s_a = 2,8,12,14,18,20,24,30$. The number of parameters, needed to specify the 3-point conformal block, besides $\alpha_{0,1,\infty}$ is $\sum_{a=1}^8 (s_a-2) =120-8$, and this is also the number of parameters one has to specify to pick out a vacuum of the 3d theory. For finite $m_s$, the number of moduli is larger. The bulk theory has $368-8 $ Coulomb moduli. The number of vacua of the 3d quiver gauge theory is now the number of positive roots with multiplicities: $32$ positive roots have multiplicity $1$; $30$ positive roots have multiplicity $2$ since they have intersection number $-1$ with $2$ of the weights; $32$ more have multiplicity $2$ as they have intersection number $-2$ with one of the weights; $44$ have multiplicity $3$ as they have intersection number $-1$ with $3$ weights; $8$ positive roots come with multiplicity $3$ because they have intersection number $-3$ with one weight; $14$ positive roots have multiplicity $4$, since they have intersection number $-1$ with $4$ weights. The net count is $368$, and we still need to subtract $8$ for the net rank constraints. 
  
\section*{Acknowledgments}
We are grateful to Tudor Dimofte, Davide Gaiotto, Ori Ganor, Andrei Okounkov, Nicolai Reshetikhin, Christian Schmidt, Shamil Shakirov, Nathan Seiberg, Cumrun Vafa and Dan Xie for helpful discussions. 
The research of M.A. and N. H. is supported in part by the Berkeley Center for Theoretical Physics, by the National Science Foundation (award number 0855653), by the Institute for the Physics and Mathematics of the Universe, and by the US Department of Energy under Contract DE-AC02-05CH11231. 
\appendix
\section{${\cal W}_{{\bf q, t}}({\bf g})$ Algebra and Vertex Operators}

Here, we will review the construction of ${\cal W}_{q,t}({\bf g})$, the $q,t$-deformed ADE  ${\cal W}$-algebra, its screening charges and vertex operators. The section it taken from \cite{FR1}, with minor changes of notation, and specializing to the simply laced case.
Consider the Heisenberg algebra, generated by modes of $n$ bosons, with relations

$$
[e_a[k], e_b[m]] = {1\over k} (q^{k\over 2} - q^{-{k\over 2}})(t^{{k\over 2} -}-t^{-{k\over 2} })C_{ab}(q^{k\over 2} , t^{k\over 2} ) \delta_{k, -m}
$$
where $C_{ab}(q,t)$ is a $q,t$ deformed Cartan matrix, $C_{ab}(q,t)= (v+v^{-1})\delta_{ab} - I_{ab}$, $I_{ab}$ is the incidence matrix of the Dynkin diagram, and  $v = \sqrt{q/t}$. The generators $e_{a}[m]$ are called the "root type" generators in \cite{FR1}. They also introduce fundamental weight type generators $w_a[m]$
$$
[e_a[k], w_a[m]] ={1\over k} (q^{k\over 2}  - q^{-{k\over 2} })(t^{{k\over 2} }-t^{-{k\over 2} })\delta_{ab}\delta_{k, -m}
$$
satisfying
\beq\label{etow}
e_a[k] = \sum_{b=1}^n C_{ab}(q^{k},t^k) w_b[k].
\eeq
The screening charge operators are 
$$
S_a(x) = : \exp\Bigl(\sum_{k\neq 0}{ e_a[k] \over q^{k\over 2} - q^{-{k \over 2}}} e^{kx}\Bigr):.
$$
(\cite{FR1} introduce another set of screening charges that correspond to D3 branes wrapping a complementary subspace of ${\mathbb R}^4$, the one rotated by $t$ instead of $q$, and which we won't need here. Furthermore, for brevity we are cavalier about the zero modes.)  In the limit where $R$ goes to zero, this becomes $S_a(x) = : \exp\Bigl(\sum_{k\neq 0}{ e_a[k] \over k\,\epsilon_1 } e^{kx}\Bigr):$, which are the usual expressions for screening charges of bosons $\varphi^{a}(x)=(e_a, \varphi(x)) =\sum_{k\in {\mathbb Z}}\frac{e_a[k]}{k}e^{kx}.$
It is easy to see that the two point functions of the screening charges exactly reproduce \eqref{sctda}, \eqref{sctdab}.

We associate primary vertex operators to a collection of $n+1$ weights $\omega_i$ such that any $n$ of them provide a basis of the weight space, and $\sum_{i=1}^{n+1} \omega_i=0$, and in addition we require $-\omega_i$ to lie in the Weyl orbit of one of the fundamental weights $w_a$. Then, the corresponding vertex operator is given by
\beq\label{cv}
:\prod_{i=1}^{n+1} V_{\omega_i}(x_i): 
\eeq
where $V_{\omega_i}$ itself is constructed out of fundamental weight and simple root vertex operators, as in section \ref{sec:Examples}.\footnote{These are closely related to the $Y$ and $A$ operators of \cite{FR1}.}
\beq\label{weightv}
W_a(x) =:  \exp\Bigl(  \;\sum_{k\neq 0}{w_a[k] \over (q^{k\over 2} - q^{-{k \over 2}})(t^{k\over 2} - t^{-{k \over 2}})} \, e^{kx}\Bigr) :
\eeq
and 
\beq\label{rootv}
E_a(x) =:  \exp\Bigl(  \;\sum_{k\neq 0}{e_a[k] \over (q^{k\over 2} - q^{-{k \over 2}})(t^{k\over 2} - t^{-{k \over 2}})}\, e^{kx}\Bigr) :
\eeq
We can relate the two sets of vertex operators, using the relation \eqref{etow}.

To take the limit back to the ${\cal W}({\bf g})$ algebra we write
$q = e^{R \epsilon_1}, t=e^{-R \epsilon_2}$ and $e^{x_i} = z e^{R \alpha_i}$, and take $R$ to zero, while rescaling $e_a[k]$ and $w_a[k]$ by a power of $R$ for the Heisenberg algebra to continue making sense. The momentum $\alpha$ carried by $V_{\alpha}(z)$ is $\alpha = \sum_{i=1}^{n+1} \alpha_i \,\omega_i$.
The individual $V_{\omega_i}(x_i)$ do not have a good conformal limit, but the products in \eqref{cv} do,  \eqref{primary}:
$$
:\prod_{i=1}^{n+1} V_{\omega_i}(x_i): \qquad \rightarrow \qquad V_{\alpha}(z).
$$

\bibliographystyle{utcaps}
\bibliography{myrefs}
\end{document}